\documentclass[journal]{IEEEtran}
\pdfoutput=1
\usepackage{epsfig,latexsym,amssymb,amsmath,graphics}
\usepackage{graphicx}
\usepackage{amssymb,dsfont,amsthm}
\usepackage{cite,color}
\usepackage{url}
\usepackage{booktabs}
\usepackage{algorithm, algorithmicx}
\usepackage{algpseudocode}
\usepackage{stfloats}
\usepackage{verbatim}
\usepackage{epstopdf}
\usepackage{subfig}
\usepackage{bm}
\usepackage{setspace}
\usepackage{siunitx}
\usepackage{float}
\definecolor{ColorRed}{rgb}{1,0,0}

\makeatletter

\newcommand{\Rmnum}[1]{\expandafter\@slowromancap\romannumeral #1@}
\makeatother

\begin{document}

\captionsetup[figure]{name={Fig. },labelsep=period}
	
\title{Ultraviolet Positioning via TDOA: Error Analysis and System Prototype}

\author{Shihui Yu, Chubing Lv, Yueke Yang, Yuchen Pan, Lei Sun, Juliang Cao, Ruihang Yu, Chen Gong, Wenqi Wu, and Zhengyuan Xu
	\thanks{This work was supported in part by the National Natural Science Foundation of China under Grant 62171428.
	
       Shihui Yu, Yueke Yang, Yuchen Pan, Lei Sun, Chen Gong, and Zhengyuan Xu are with Key Laboratory of Wireless-Optical Communications, Chinese Academy of Sciences, University of Science and Technology of China, Hefei, Anhui 230027, China. Email addresses: \{zkd708843181, ykyang1106, ycpan, sunlei1008\}@mail.ustc.edu.cn, \{cgong821, xuzy\}@ustc.edu.cn. 
       
       Chubing Lv, Juliang Cao, Ruihang Yu, and Wenqi Wu are with College of Intelligence Science and Technology, National University of Defense Technology, Changsha, Hunan 410073, China. Email addresses: \{lvchubing13, jlcao, yuruihang\}@nudt.edu.cn, wenqiwu\_lit@hotmail.com.
}
%	\thanks{This work was supported by Key Program of National Natural Science Foundation of China (Grant No. 61631018),
%		Key Research Program of Frontier Sciences of CAS (Grant No. QYZDY-SSW-JSC003),
%		and the Fundamental Research Funds for the Central Universities.
%		
%		Yuan Wang, Chen Gong, and Zhengyuan Xu are with Key Laboratory of Wireless-Optical Communications, Chinese Academy of Sciences, University of Science and Technology of China, Hefei, Anhui 230027, China.
%		Email: wangy001@mail.ustc.edu.cn; \{cgong821,xuzy\}@ustc.edu.cn.
%		
%		}}
}
\maketitle
\vspace{-1.6cm}  %1.6cm即数值，自行调节即可。

\begin{abstract}
	
	This work performs the design, real-time hardware realization, and experimental evaluation of a positioning system by ultra-violet (UV) communication under photon-level signal detection. The positioning is based on time-difference of arrival (TDOA) principle. Time division-based transmission of synchronization sequence from three transmitters with known positions is applied. We investigate the positioning error via decomposing it into two parts, the transmitter-side timing error and the receiver-side synchronization error. The theoretical average error matches well with the simulation results, which indicates that theoretical fitting can provide reliable guidance and prediction for hardware experiments. We also conduct real-time hardware realization of the TDOA-based positioning system using Field Programmable Gate Array (FPGA), which is experimentally evaluated via outdoor experiments. Experimental results match well with the theoretical and simulation results.

\end{abstract}


\section{Introductions}
%\subsection{Background}
%Wireless optical communication (free space optical communication) technology, compared with the traditional radio communication, effectively avoid the problem of insufficient the channel capacity, insufficient spectrum resources and serious electromagnetic interference; Compared with optical fiber communication, it can also cope with the problem that wired optical communication system is not free enough and has poor mobility. Among them, ultraviolet communication not only has the above advantages of wireless optical communication, compared with other bands of optical communication, ultraviolet light due to its own scattering characteristics, carrier propagation path is more diverse, so ultraviolet communication has been widely concerned.
Terminal positioning has a wide variety of applications, relying on frequency, time and spatial information. Timing-based positioning techniques rely on the measurements of flying times of signals between nodes. Among them, time difference of arrivals (TDOA) techniques are effective in locating a source based on intersections of hyperbolic curves defined by the arrival time differences of received signal and are further improved with bias reduction \cite{301830,6151186}. Solution and performance analysis of geolocation by TDOA are investigated \cite{259534,4682575}. Work \cite{5778025} develops an linear solution that estimate the target location based on TDOA technique for ultra-wide band (UWB) localization. Work \cite{1323254} proposes an algebraic solution for the position and velocity of a moving terminal using TDOA of signals received at a number of receivers. Geolocation based on biased range estimates is proposed \cite{6151850}. A TDOA-based optical wireless indoor localization approach is proposed using light emitting diodes (LEDs) and the performance is demonstrated \cite{6131130}.
%Due to the scattering of ultraviolet (UV) signals in the atmosphere, UV communication can be applied to the scenario where the transmitter and receiver cannot be perfectly aligned.

On the other hand, optical wireless communication (OWC) refers to transmission in unguided propagation media through optical carriers, i.e., visible, infrared (IR) and ultra-violet (UV) band \cite{6844864}. In comparison to radio-frequency (RF) counterparts, the OWC link shows potential immunity to electromagnetic interference and does not have electromagnetic radiation \cite{1159099,5175684,6165319,6458971}. It also provides a high reuse factor, an inherent security, and robustness to electromagnetic interference \cite{4063386}. Laser-based free-space optical (FSO) systems are also appealing for a wide range of applications, for example, an envisioned campus connectivity scenario where inter-building connections are enabled by high data rate FSO links or high quality video surveillance. The monitoring of a city can be made possible by FSO links \cite{4063386,938713,1299334}. The UV-C band is essentially solar blind, where most solar radiation is getting absorbed by the ozone layer in the upper atmosphere, which results in an almost noiseless transmission channel \cite{1527982}. Hence, a photon detector can approach quantum-limited photon counting detection performance \cite{8641355}. Besides, line-of-sight (LOS) configurations in the UV-C band have additional desirable characteristics compared to conventional optical wireless networks operating outside this band.

Due to almost negligible background radiation in the UV spectrum close to the earth surface, high-sensitivity photon-level detection can be developped based on multi-stage amplification \cite{1527982,10.1117/12.582002}, for example using a photomultiplier tube (PMT). The channel characteristics of UV communication have been demonstrated experimentally in \cite{5342313,Xiao2011NonlineofsightUS,6923957,Liao2015LongdistanceNU,Borah2021SingleAD,9295804}, where Monte-Carlo approach is adopted to trace the transmission path of a photon experiencing random scattering and attenuation events. Relay communication protocols have been designed and studied in \cite{6877718,8275030,7830279}. In \cite{5599260}, generalized maximum likelihood sequence detection has been investigated. To increase the detection sensitivity, work \cite{7112175,7047703} studied the signal detection with receiver diversity. Experimental work on long-distance channel is introduced in \cite{6923957}. In \cite{8332484}, the throughput of 1Mbps has been demonstrated via receiver diversity over 1 km. Work \cite{2022arXiv220801559Y} proposes an upper bound on the synchronization mean squared error of an OWC system with a photon-counting receiver.

In recent years, the application of optical wireless transmission for positioning has attracted much attention. Various algorithms and systems for visible light positioning (VLP) have been proposed \cite{8322671,8121297,7394267,8533834,7859314}. For example, a positioning system for underwater terminals is proposed based on the underwater wireless optical code division multiple access (CDMA) network \cite{7928991}. In work \cite{10092387}, a full-duplex asymmetric underwater optical wireless communication and positioning (UOWC-UOWP) system for future underwater wireless sensor network (UWSN) is proposed. Researchers propose a visible light communication (VLC) -based ID-estimation and LED-tracking method for indoor positioning contexts \cite{7275471}. For UV band, researchers present a probabilistic fire location estimation in indoor fire environments by fusing two rotary UV sensors \cite{9158795}. However, terminal positioning based on UV transmission needs to be explored.

This work aims to design a positioning system by UV transmission under photon-level signal detection, assuming on-off keying (OOK) modulation using UV LEDs. The positioning is based on TDOA principle. We utilize the coordinates of the known transmitters to locate the unknown receiver, and time division transmission of synchronization sequence is applied. Our work investigates the positioning error via decomposing it into two parts. The theoretical average error matches well with the simulation results. We also conduct real-time hardware realization of the TDOA-based positioning system using Field Programmable Gate Array (FPGA), which is evaluated via outdoor real-time experiments. Experimental results match well with the theoretical and simulation results. The positioning accuracy can be further improved when the clock alignment error can be reduced at the transmitters.

%\subsection{Organization}
The remainder of this paper is organized as follows. In Section \ref{sec.system_model}, the positioning model and TDOA principle are introduced. In Section \ref{Transmittion scheme}, we decompose the timing error into two parts, the transmitter-side synchronization error and receiver-side synchronization error, and analyze each part. In Section \ref{Theoretical fitting and simulation verification}, simulation results are obtained to verify the theoretical average positioning error. In Section \ref{Outdoor Location Experiment}, we introduce the real-time hardware realization, as well as the experimental results in two outdoor real-time experiments on the university playground and a larger lawn, where the experimental results match well with the theoretical average error. Finally, Section \ref{section.conclusion} concludes this work.

\section{Positioning Model} \label{sec.system_model}
The arrival time difference of the signal can be accurately recorded through synchronization process, based on which the terminal positioning can be achieved.
%\textcolor{ColorBlack}{
	\subsection{The TDOA-Based Positioning Protocol}
\begin{figure*}[htbp]
		\centering
		\includegraphics[width=4.5in]{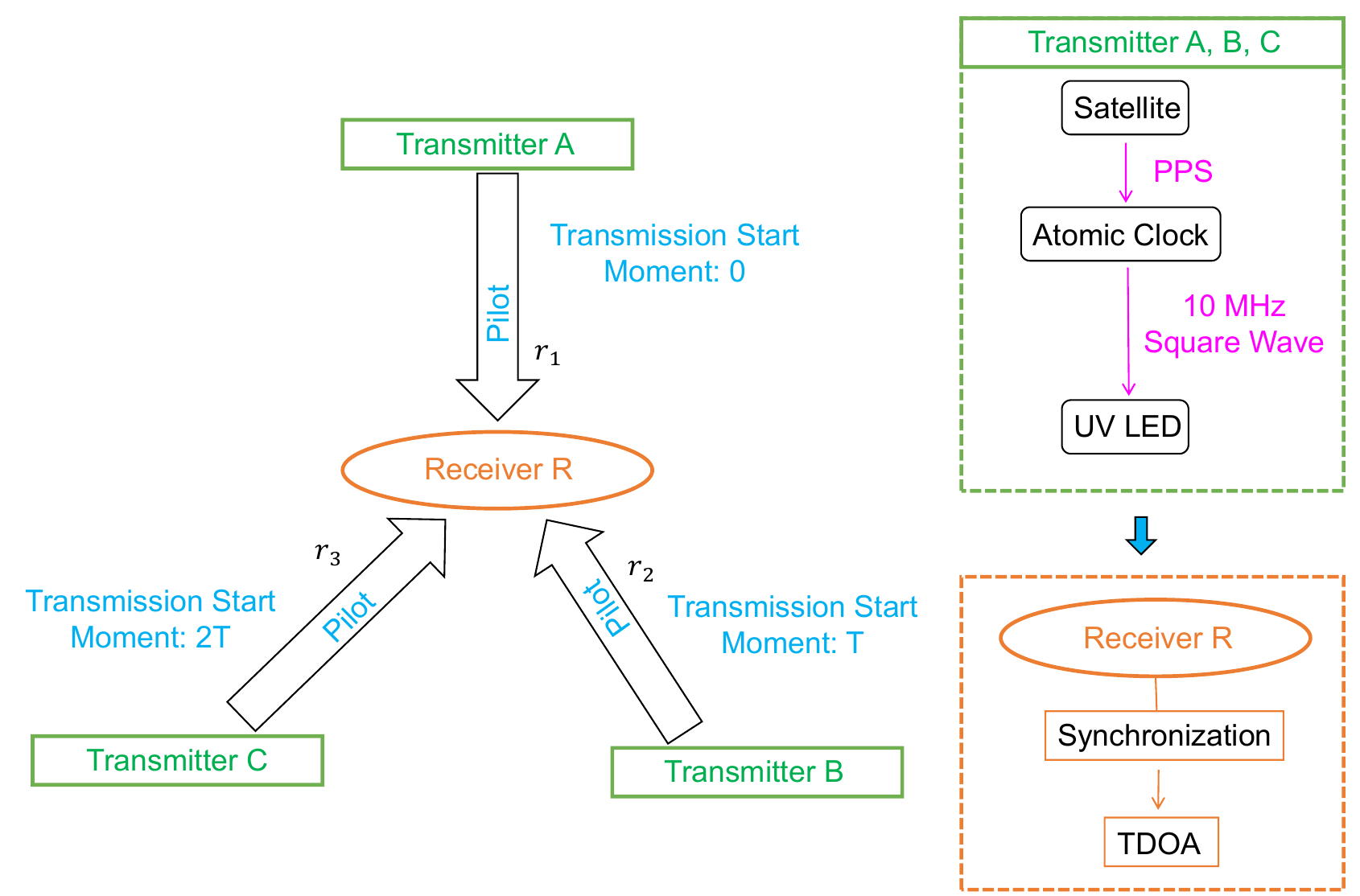}
		\caption{$\ $Schematic diagram of TDOA principle.} \label{TDOA}
		%    	\vspace{-1.0cm}%%压缩图片后间隔
	\end{figure*}

	Consider 2-dimensional positioning, while 3-dimensional positioning can be derived in a similar fashion. Assume that transmitters A, B and C are three separate anchor points, which send signals to receiver R, as shown in Fig. \ref{TDOA}. The three transmitters are aligned via an unified clock in the form of pulse per second (PPS) from a satellite timing system. The transmitters send synchronization signals to the receiver under an unified clock in a time-division manner at fixed time interval $T$. The time-division transmission is controlled by $10MHz$ square wave generated by an atomic clock drived by the PPS signal. Based on the synchronization of pilots from the three transmitters, the receiver can obtain the time difference of arrival, from which the receiver location can be estimated.
	
	Assume that the transmitters send synchronization signals via OOK modulation. Due to extremely weak light intensity, the received signal can be characterized by discrete photoelectrons, and the number of detected photoelectrons during a certain duration follows Poisson distribution. To remove the inter-transmitter interference, assume that $T$ is longer than the synchronization length, such that the signals received by the receiver from different transimitters will not overlap. Let $\lambda_{s, A}$ and $\lambda_b$ denote the mean numbers of detected photoelectrons for the signal from transmitter A and background radiation components, respectively. Assume that the three transmitters adopt the same synchronization sequence $\{s_i, 1 \leq i \leq L\}$, where $L$ is the synchronization sequence length. Let $s_i$ denote the $i^{th}$ symbol of pilot sequence and $N_{i,A}$ denote the number of received photoelectrons from the $i^{th}$ symbol, satisfying
	\begin{align} \label{equ.Poisson_channel}
		P(N_{i,A} = u \mid s_i=1) &= \frac{(\lambda_{s, A}+\lambda_b)^{u}}{{u}!}e^{-(\lambda_{s,A}+\lambda_b)},\\
		P(N_{i,A} = u \mid s_i=0) &= \frac{\lambda_b^{u}}{{u}!}e^{-\lambda_b}.
	\end{align} 
	For transmitter B, we have
	\begin{align} \label{equ.Poisson_channel2}
		P(N_{i,B} = u \mid s_i=1) &= \frac{(\lambda_{s, B}+\lambda_b)^{u}}{{u}!}e^{-(\lambda_{s,B}+\lambda_b)},\\
		P(N_{i,B} = u \mid s_i=0) &= \frac{\lambda_b^{u}}{{u}!}e^{-\lambda_b}.
	\end{align} 
	Similarly for transmitter C, we have
	\begin{align} \label{equ.Poisson_channel3}
		P(N_{i,C} = u \mid s_i=1) &= \frac{(\lambda_{s, C}+\lambda_b)^{u}}{{u}!}e^{-(\lambda_{s,C}+\lambda_b)},\\
		P(N_{i,C} = u \mid s_i=0) &= \frac{\lambda_b^{u}}{{u}!}e^{-\lambda_b}.
	\end{align} 
	
	Let $T_s$ denote the symbol duration. In this work, each symbol duration is divided into $n$ chips with duration $T_c = \frac{T_s}{n}$. For a certain chip index $t$, the number of pulses in that chip is denoted as $ N_{t,A} $. Let ${\hat t}_{Astart}$ denote the starting chip index estimate of the synchronization sequence. The synchronization solution is formulated by finding the following start index
	\begin{equation}
		\begin{aligned} \label{equ.auto-correlation}
			{\hat t}_{Astart}&=\arg \max _{t_{Astart}}\left\{\sum_{i=1}^{L}\left\{u_{i,A}^{t_{Astart}} \times\left(2 s_{i}-1\right)\right\}\right\},\\
		\end{aligned}
	\end{equation}
	where
	\begin{equation}
		\begin{aligned} \label{equ.n}
			u_{i,A}^{t_{Astart}}&=\sum_{j=0}^{n-1}N_{t_{Astart}+n\times (i-1) +j,A}~,
		\end{aligned}
	\end{equation}
	denotes the number of pulses correlated with symbol $ s_{i} $ based on start chip index $ t_{Astart} $, and symbol $ s_{i} $ is transformed to $ \left(2 s_{i}-1\right) $ belonging to $ \{-1,1\} $ for correlation. Maximum correlation peak criterion is adopted. $ \hat{t}_{Astart} $ is regarded as the time when the synchronization signal reaches the receiver. Similarly, we have
	\begin{equation}
		\begin{aligned} \label{equ.auto-correlation2}
			{\hat t}_{Bstart}&=\arg \max _{t_{Bstart}}\left\{\sum_{i=1}^{L}\left\{u_{i,B}^{t_{Bstart}} \times\left(2 s_{i}-1\right)\right\}\right\},\\
		\end{aligned}
	\end{equation}
	\begin{equation}
		\begin{aligned} \label{equ.auto-correlation3}
			{\hat t}_{Cstart}&=\arg \max _{t_{Cstart}}\left\{\sum_{i=1}^{L}\left\{u_{i,C}^{t_{Cstart}} \times\left(2 s_{i}-1\right)\right\}\right\}.\\
		\end{aligned}
	\end{equation}
	
	\subsection{Positioning Algorithm: TDOA}
	TDOA measures the arrival time difference of signals from the transmitters.    
	Assume that the arrival time instants of the synchronization signal from the three transmitters are ${t}_A$, ${t}_B$ and ${t}_C$. The difference of flying time can be obtained accordingly considering the transmission instants from the three transmitters. Let $t_{BA}$ denote the flying time difference between signals from B and A, and $t_{CB}$ denotes the flying time difference between signals from C and B. The positioning is based on the following property
	
	%    \begin{figure}[htbp]
		%    	\centering
		%    	\includegraphics[width=2.5in]{TDOA}
		%    	\caption{$\ $Schematic diagram of TDOA principle.} \label{TDOA}
		%%    	\vspace{-1.0cm}%%压缩图片后间隔
		%    \end{figure}
	
	\begin{equation}\label{d_diff}
		\begin{aligned} 
			r_{21}&=r_2-r_1=c \times t_{BA},\\
			r_{32}&=r_3-r_2=c \times t_{CB},
		\end{aligned}
	\end{equation}
	where $c$ is the speed of light.
	
	Assume that the coordinates of transmitters A, B and C are $ \left(x_A,y_A\right),~ \left(x_B,y_B\right)$ and $ \left(x_C,y_C\right)$, respectively. The coordinate of R, denoted as $\left(x,y\right)$, can be obtained via solving Eq. \eqref{lequ} .
	\begin{figure*}[hb] % hb底部，ht为头部
		\centering % 公式居中
		\hrulefill % 添加一条水平线
		\vspace*{8pt} % 调整线与公式之间的距离
		\begin{equation}
			\begin{aligned} \label{lequ}
				\left\{
				\begin{array}{lr}
					r_{21}&=\sqrt{\left(x_B-x\right)^2+\left(y_B-y\right)^2}-\sqrt{\left(x_A-x\right)^2+\left(y_A-y\right)^2},\\
					r_{32}&=\sqrt{\left(x_C-x\right)^2+\left(y_C-y\right)^2}-\sqrt{\left(x_B-x\right)^2+\left(y_B-y\right)^2}.
				\end{array}
				\right.
			\end{aligned}	
		\end{equation}
		%\begin{equation}
		%		\begin{aligned}	 \label{lequ2}
			%			r_{32}=\sqrt{\left(x_C-x\right)^2+\left(y_C-y\right)^2}-\sqrt{\left(x_B-x\right)^2+\left(y_B-y\right)^2}.
			%		\end{aligned}
		%%		\end{array}
	%%		\right.
	%\end{equation}
\end{figure*}
\subsection{Positioning Mean Squared Error (MSE)}\label{MSE Calculation}
The positioning MSE can be approximated as follows. Denote the true receiver location as $\left(x_0,y_0\right)$. Let error vector $\Delta \boldsymbol{u} = \left[\Delta x, \Delta y\right]^T=\left[x-x_0, y-y_0\right]^T$, and $\Delta r_{21}$ and $\Delta r_{32}$ denote the gap between estimated value and actual distance difference
\begin{equation}\label{tba}
	\begin{aligned}
		\Delta r_{21}=\hat{r}_{21}-r_{21}=c\times \Delta t_{BA},
	\end{aligned}
\end{equation}
\begin{equation}\label{tcb}
	\begin{aligned}
		\Delta r_{32}=\hat{r}_{32}-r_{32}=c\times \Delta t_{CB},
	\end{aligned}
\end{equation}
where $\Delta t_{BA}$ denotes the estimation error of $t_{BA}$ and $\Delta t_{CB}$ denotes the estimation error of $t_{CB}$. Distance differences $\Delta r_{21}$ and $\Delta r_{32}$ can be estimated via the first-order partial differential
\begin{equation}
	\begin{aligned}
		\Delta r_{21}=\frac{\partial r_{21}}{\partial x}\big|_{x=x_0} \times \Delta x +\frac{\partial r_{21}}{\partial y}\big|_{y=y_0} \times \Delta y,
	\end{aligned}
\end{equation}
\begin{equation}
	\begin{aligned}
		\Delta r_{32}=\frac{\partial r_{32}}{\partial x}\big|_{x=x_0} \times \Delta x +\frac{\partial r_{32}}{\partial y}\big|_{y=y_0} \times \Delta y,
	\end{aligned}
\end{equation}
where the higher-order terms are ignored.
Thus, we have
\begin{equation}
	\begin{aligned} 
		\begin{bmatrix} \frac{\partial r_{21}}{\partial x}\big|_{x=x_0} & \frac{\partial r_{21}}{\partial y}\big|_{y=y_0} \\ \frac{\partial r_{32}}{\partial x}\big|_{x=x_0} & \frac{\partial r_{32}}{\partial y}\big|_{y=y_0} \end{bmatrix} \begin{bmatrix}
			\Delta x\\
			\Delta y
		\end{bmatrix}=
		\begin{bmatrix}
			c\times \Delta t_{BA}\\
			c\times \Delta t_{CB}
		\end{bmatrix}.
	\end{aligned}
\end{equation}
From Eq. \eqref{lequ}, we have Eq. \eqref{matequ}.
%\end{figure*}
\begin{figure*}[hb] % hb底部，ht为头部
	\centering % 公式居中
	\hrulefill % 添加一条水平线
	\vspace*{8pt} % 调整线与公式之间的距离
	\begin{equation}\label{matequ}
		\begin{aligned} 
			\begin{bmatrix} \frac{x_A-x_0}{r_1}-\frac{x_B-x_0}{r_2} & \frac{y_A-y_0}{r_1}-\frac{y_B-y_0}{r_2} \\ \frac{x_B-x_0}{r_2}-\frac{x_C-x_0}{r_3} & \frac{y_B-y_0}{r_2}-\frac{y_C-y_0}{r_3} \end{bmatrix} \begin{bmatrix}
				\Delta x\\
				\Delta y
			\end{bmatrix}=
			\begin{bmatrix}
				c\times \Delta t_{BA}\\
				c\times \Delta t_{CB}
			\end{bmatrix}.
		\end{aligned}
	\end{equation}
\end{figure*}

Denote the $2\times2$ matrix in Eq. \eqref{matequ} as $\boldsymbol{G}$ and the vector in the RHS as $\boldsymbol{h}$. When $\boldsymbol{G}$ is full rank, we have $\Delta \boldsymbol{u} = \boldsymbol{G}^{-1}\boldsymbol{h}$. Thus, the positioning MSE matrix is given as follows
\begin{equation}\label{mseforlocation}
	\boldmath{E}\left[\Delta u \Delta u^T\right]=G^{-1}\boldmath{E}\left[hh^T\right]G^{-T},
\end{equation}
where
\begin{equation}\label{mseforlocation2}
	\boldsymbol{h} \boldsymbol{h}^T=\left[\begin{array}{cc}
		\Delta \mathrm{r}_{21}^2 & \Delta \mathrm{r}_{21} \Delta \mathrm{r}_{32} \\
		\Delta \mathrm{r}_{32} \Delta \mathrm{r}_{21} & \Delta \mathrm{r}_{21}^2
	\end{array}\right].
\end{equation}
Letting $\sigma^2_A$, $\sigma^2_B$ and $\sigma^2_C$ denote the estimate MSE of flying time $t_A$,  $t_B$ and $t_C$, respectively, we have
\begin{equation}\label{mseforlocation3}
	\frac{\boldsymbol{E}\left(\boldsymbol{h} \boldsymbol{h}^T\right)}{c^2}=\left[\begin{array}{cc}
		\sigma^2_A+\sigma^2_B & -\sigma^2_B \\
		-\sigma^2_B & \sigma^2_B+\sigma^2_C
	\end{array}\right].
\end{equation}
The positioning error can be expressed as the square root of the trace of MSE matrix as follows
\begin{equation}\label{trace}
	e_p=\sqrt{tr(\boldsymbol{E}\left[\boldsymbol{\Delta u \Delta u^T}\right])}.
\end{equation}

%\vspace{0.3cm}

%\graphicspath{{figures/Fig_PowerAlloc/}}
%\graphicspath{{figure_yu/FigSync/}}

\section{Theoretical fitting principle for positioning error} \label{Transmittion scheme}
We first investigate the decomposition of estimation error of flying time $t_A$, $t_B$ and $t_C$ into two parts, and analyze each part. 

\subsection{Estimation Error Decomposition for the Flying Time}\label{fly}
The positioning error primarily consists of two parts, one from the atomic clock timing at the transmitter side, and the other from the synchronization error at the receiver side. 

Since it is difficult to achieve perfect alignment of the rising edge from different transmitter-side clocks, the rising edge difference of $10MHz$ square wave multiplied by the speed of light leads to a large distance error. Such error will be characterized in Section \ref{timesec} in detail. Besides, synchronization also leads to estimation error of the correlation peak. An upper bound on the synchronization MSE will be analyzed in Section \ref{syncerror} in detail.

Let $\sigma^2_{clock}$ denote the rising edge error of the atomic clock, and $\sigma^2_{syncA}$, $\sigma^2_{syncB}$ and $\sigma^2_{syncC}$ denote the synchronization error of synchronization signal from transmitter A, transmitter B and transmitter C, respectively. Since the rising edge error of the transmitter clock is independent of the synchronization process, we can approximate $\sigma^2_A=\sigma_{clock}^2+\bar{\sigma}^2_{syncA}$, similarly for $\sigma^2_B$ and $\sigma^2_C$. The positioning error can be obtained from Eq. \eqref{mseforlocation}, Eq. \eqref{mseforlocation2}, Eq. \eqref{mseforlocation3} and Eq. \eqref{trace}.

\subsection{Timing Error of Atomic Clock}\label{timesec}
\begin{figure*}[htpb]
	\centering
	\includegraphics[width=4.5in]{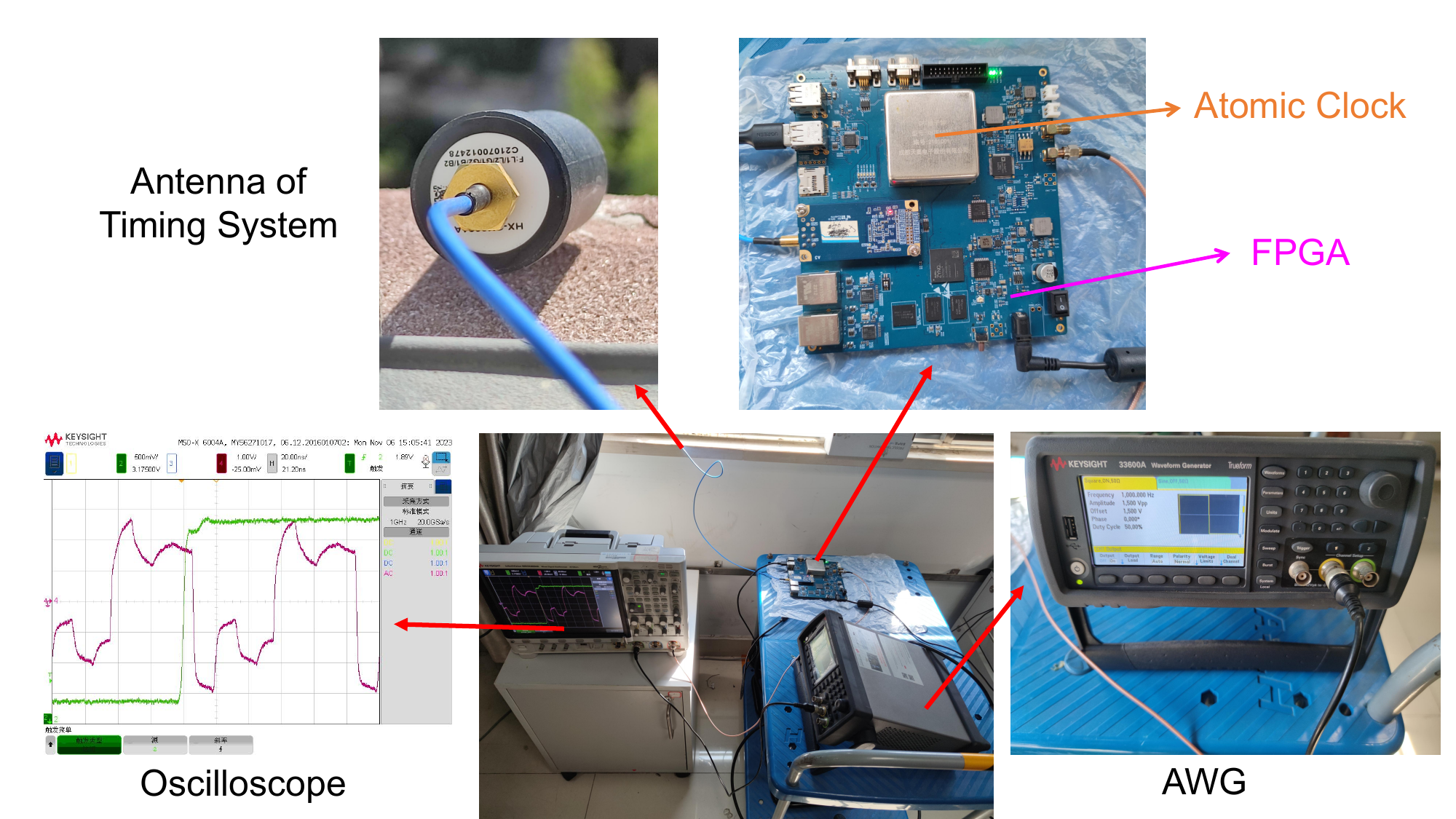}
	\caption{$\ $Experimental platform of the atomic clock timing error characterization.} \label{clockscene}
	%	\vspace{-0.5cm}%%压缩图片后间隔
\end{figure*}
The transmitters are calibrated by the satellite timing module each second, which is achieved by an atomic clock controlling the time division transmission. The atomic clock timing MSE is denoted as $\sigma_{clock}^2$. To explore its statistical properties, we test the rising edge error of atomic clock. Experimental platform is shown in Fig. \ref{clockscene}. A timing system with an antenna receives the satellite signal, drives the atomic clock to transmit the $10MHz$ square waves, which are then sampled by an oscilloscope. The PPS output of an any wave-form generator (AWG) is adopted as a baseline, whose interval from the next rising edge of $10MHz$ square waves is characterized as the atomic clock delay. Such delay characterizes the systematic error of the rising edge.

\begin{figure}[htpb]
	\centering
	\includegraphics[width=3.0in]{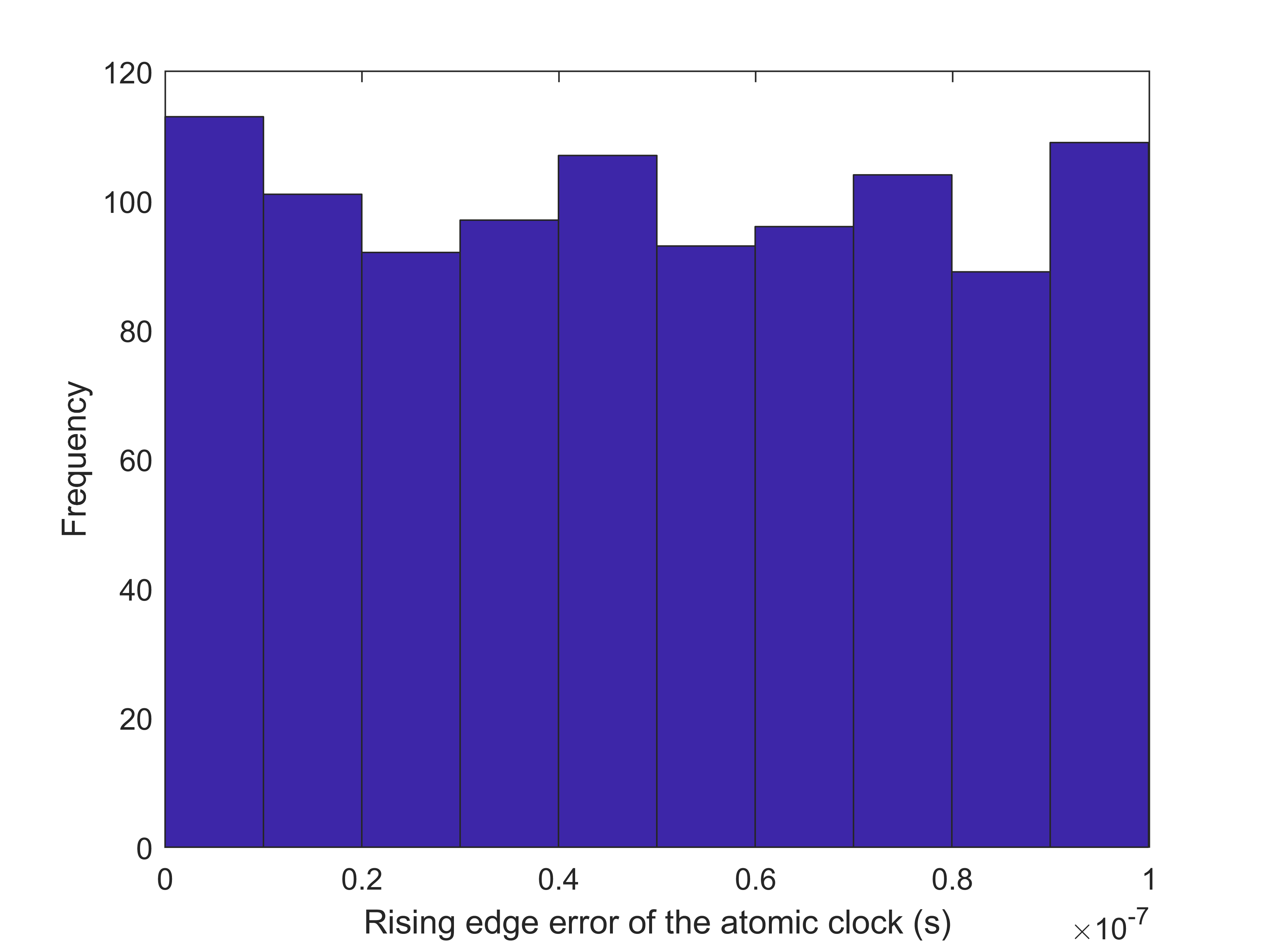}
	\caption{$\ $Empirical probability distribution of rising edge error of atomic clock.} \label{jitter}
	%	\vspace{-0.5cm}%%压缩图片后间隔
\end{figure}
The empirical probability distribution is shown in Fig. \ref{jitter}. The unit of time on the horizontal axis is one second. According to the sampling results, the atomic clock timing error can be considered to be uniformly distributed satisfying $U[0ns,100ns]$.
\subsection{Error of the Synchronization Process}\label{syncerror}
\begin{figure}[htpb]
	\centering
	\includegraphics[width=3.0in]{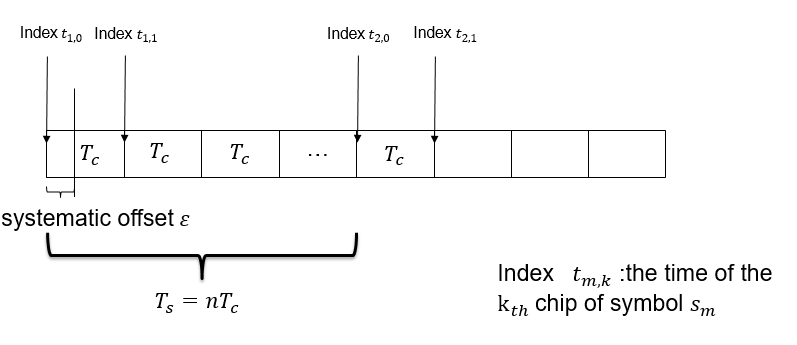}
	\caption{$\ $The timeline of the synchronization.} \label{timeline}
	%	\vspace{-0.5cm}%%压缩图片后间隔
\end{figure}

Based on our previous results on the synchronization in \cite{2022arXiv220801559Y}, the synchronization timeline is visualized in Fig. \ref{timeline}.
The process of dividing time into chips introduces systematic offsets, denoted as $ \varepsilon $ satisfying uniform distribution $ U(-\frac{T_c}{2},\frac{T_c}{2}) $.

The synchronization process can be divided into three cases. Let estimate ${\hat t}_{Astart}=t_{0,0} $ denote the accurate synchronization estimate as Case 1. In Case 2, the offset lies within a symbol's duration, i.e. ${\hat t}_{Astart}=t_{0,k} $ for certain $ 1\leq \vert k \vert \leq n-1 $, which means that the offset is $k$ chips. In Case 3, consider the offset over a symbol duration, ${\hat t}_{Astart}=t_{2m,k} $ where $  m  \geq 1  $ with $ -n \leq  k  \leq n-1 $ (or $  m  \leq -1  $ with $ -(n-1) \leq  k  \leq n $). The probability of event ${\hat t}_{Astart}=t_{m, k}$ is denoted as $p^A_{m,k}$. The upper bounds on $p^A_{m,k}$ are given by Eq. \eqref{p00}, Eq. \eqref{p0k} and Eq. \eqref{pmk}
\begin{equation}\label{p00}
	{\bar p}^A_{0, 0}=1-{\bar p}^A_{0, 1},
\end{equation}
\begin{align}\label{p0k}
	p^A_{0, k} \leq {\bar p}^A_{0, k}=\Phi\left(\frac{\frac{\lambda_{s,A}}{2}\left(2\frac{\varepsilon}{T_s}-\frac{k}{n} \right)\left(L-1\right)}{\sqrt{2\frac{k}{n} \left(\frac{\lambda_{s,A}}{2}+\lambda_{b}\right)\left(L-1\right)+\frac{\varepsilon}{T_s}\lambda_{s,A}}}\right).
\end{align}
\begin{figure*}[hb] % hb底部，ht为头部
	\centering % 公式居中
	\hrulefill % 添加一条水平线
	\vspace*{8pt} % 调整线与公式之间的距离
	\begin{equation}\label{pmk}
		\begin{aligned}
			p^A_{2m, k} \leq {\bar p}^A_{2m,k}=\Phi\left(\frac{\left(-2m\right)\frac{\lambda_{s,A}}{2}\left[1-2\left(\frac{k}{n}-\frac{\varepsilon}{T_s}\right)\right]-L\frac{\lambda_{s,A}}{2}\left(1-\frac{\varepsilon}{T_s}\right)-\frac{\lambda_{s,A}}{2}\left(2\frac{\varepsilon}{T_s}-\frac{k}{n}\right)}{\sqrt{2\left(\frac{\lambda_{s,A}}{2}+\lambda_{b}\right)\left[\left(L-m\right)-\left(-2m\right)\left(1-2\frac{k}{n}\right)-\frac{k}{n}\right]+\frac{\varepsilon}{T_s}\lambda_{s,A}}}\right).
		\end{aligned}
	\end{equation}
\end{figure*}
For estimate ${\hat t}_{Astart}=t_{m,k}$, let $e_{m,k}= t_{m,k} - \varepsilon $ denote the synchronization error. An upper bound on the MSE is denoted as $\bar{\sigma}_{syncA}^2$ in Eq. \eqref{uppbound}, 
\begin{figure*}[hb] % hb底部，ht为头部
	\centering % 公式居中
	\hrulefill % 添加一条水平线
	\vspace*{8pt} % 调整线与公式之间的距离
	\begin{align}\label{uppbound}
		\bar{\sigma}_{syncA}^2 \triangleq\int_{-\frac{T_c}{2}}^{\frac{T_c}{2}}\left[e_{0,0}^{2}  \bar{p}^A_{0,0}+2 \sum_{k=1}^{n-1} e_{0, k}^{2}  \bar{p}^A_{0, k}
		+2 \sum_{m=1}^{\infty} \sum_{k=-n}^{(n-1)} e_{2 m, k}^{2}  \bar{p}^A_{2 m, k}\right] p(\varepsilon) d \varepsilon,
	\end{align}
\end{figure*}
where $T_c$ is the chip duration, $ p\left(\varepsilon\right)= \frac{1}{T_c}$ for $ \varepsilon \in \left[-\frac{T_c}{2},\frac{T_c}{2}\right] $. We can obtain $\bar{\sigma}_{syncB}^2$ and $\bar{\sigma}_{syncC}^2$ similarly.

\section{Simulation verification for positioning error}  \label{Theoretical fitting and simulation verification}
\begin{figure}[htbp]
	\centering
	\includegraphics[width=3.0in]{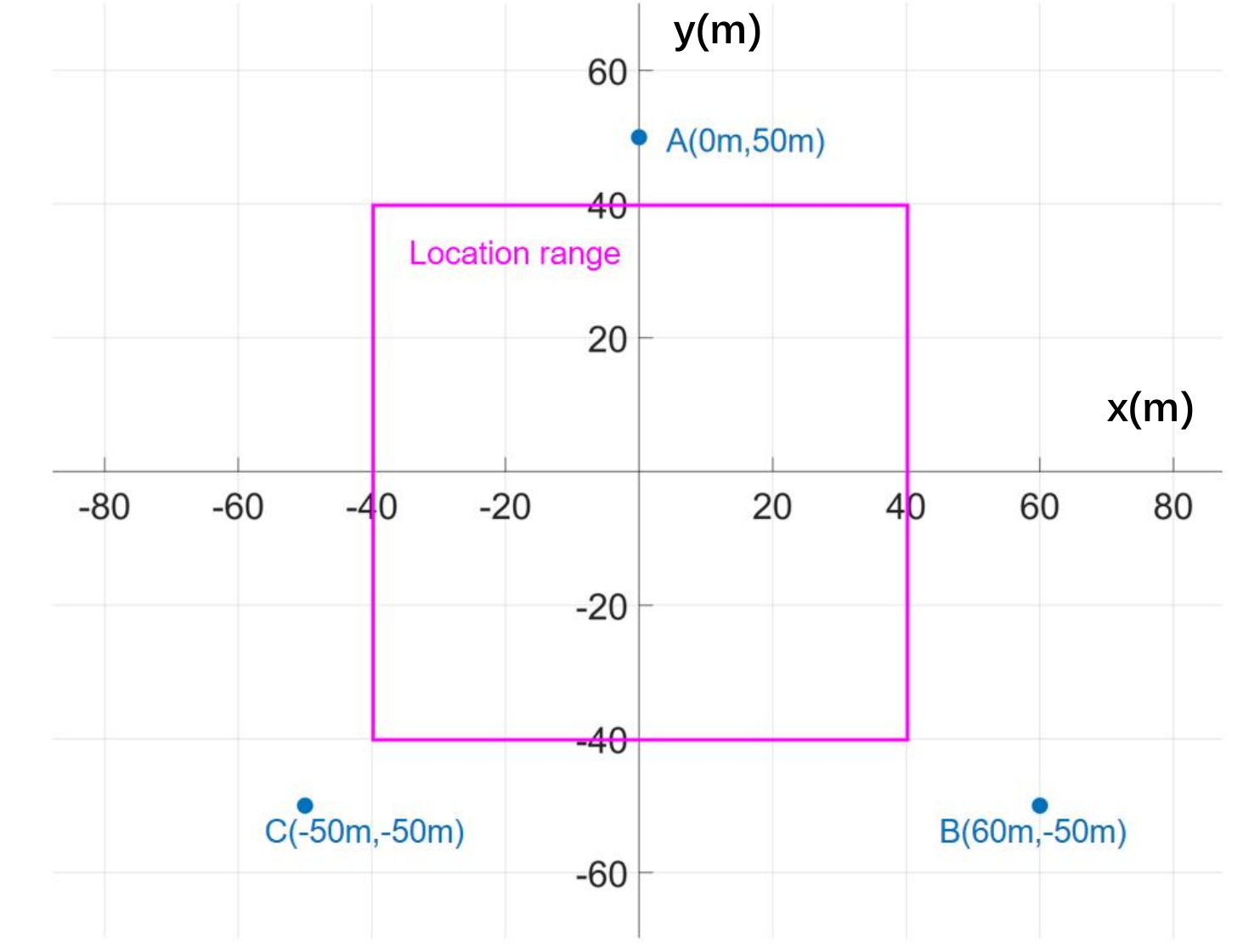}
	\caption{$\ $Schematic diagram of location range.} \label{LocationRange}
	%    	\vspace{-1.0cm}%%压缩图片后间隔
\end{figure}

We conduct simulation to show the accuracy of theoretical error as shown in Section \ref{fly}. Lay the three transmitters as shown in Fig. \ref{LocationRange}, and locate receiver R in the range represented by the purple box.

%\subsection{Time Division Transmission}\label{Time Division Transmission4}
Assume that transmitters A, B and C send the same synchronization sequence with length $L=256$ in a time-division manner with duration $T=300us$. Assume that the transmission symbol rate is $ 1Mbps$. The transmission power is $100mW$ and the detector receiving area is $1.77\times10^{-4}~m^2$. The transmitter divergence angle is $120^\circ$. Set $ \lambda_{b}=1$ as the noise intensity. After calculating the link gain for line of sight (LOS) UV communication and signal intensity $\lambda_{s}$, we can obtain the approximate upper bound on the average positioning error for each receiver position as well as the correspounding simulation results. The average error is $\sqrt{MSE}$. The sampling rate at the receiver is set to $100MHz$ due to UV PMT dead time of approximately $10$ ns. As $ \lambda_{b}$ is relatively small, we apply a cut-off function for signal intensity $\lambda_{s}$, which is clipped to $100$ if exceeding $100$.
%\begin{equation}\label{cutoff}
%	\begin{aligned} 
	%	\lambda_{s}=	\left\{
	%		\begin{array}{lr}
		%			\lambda_{s},&~~~\lambda_{s}\leq 100;\\
		%			100,&~~~\lambda_{s}\textgreater  100.
		%		\end{array}
	%		\right.
	%	\end{aligned}
%\end{equation}

\subsection{Ideal Case: Perfectly Synchronized Transmitters}\label{ssec.ideal}
\begin{figure}[htbp]
	\centering
	\includegraphics[width=3.0in]{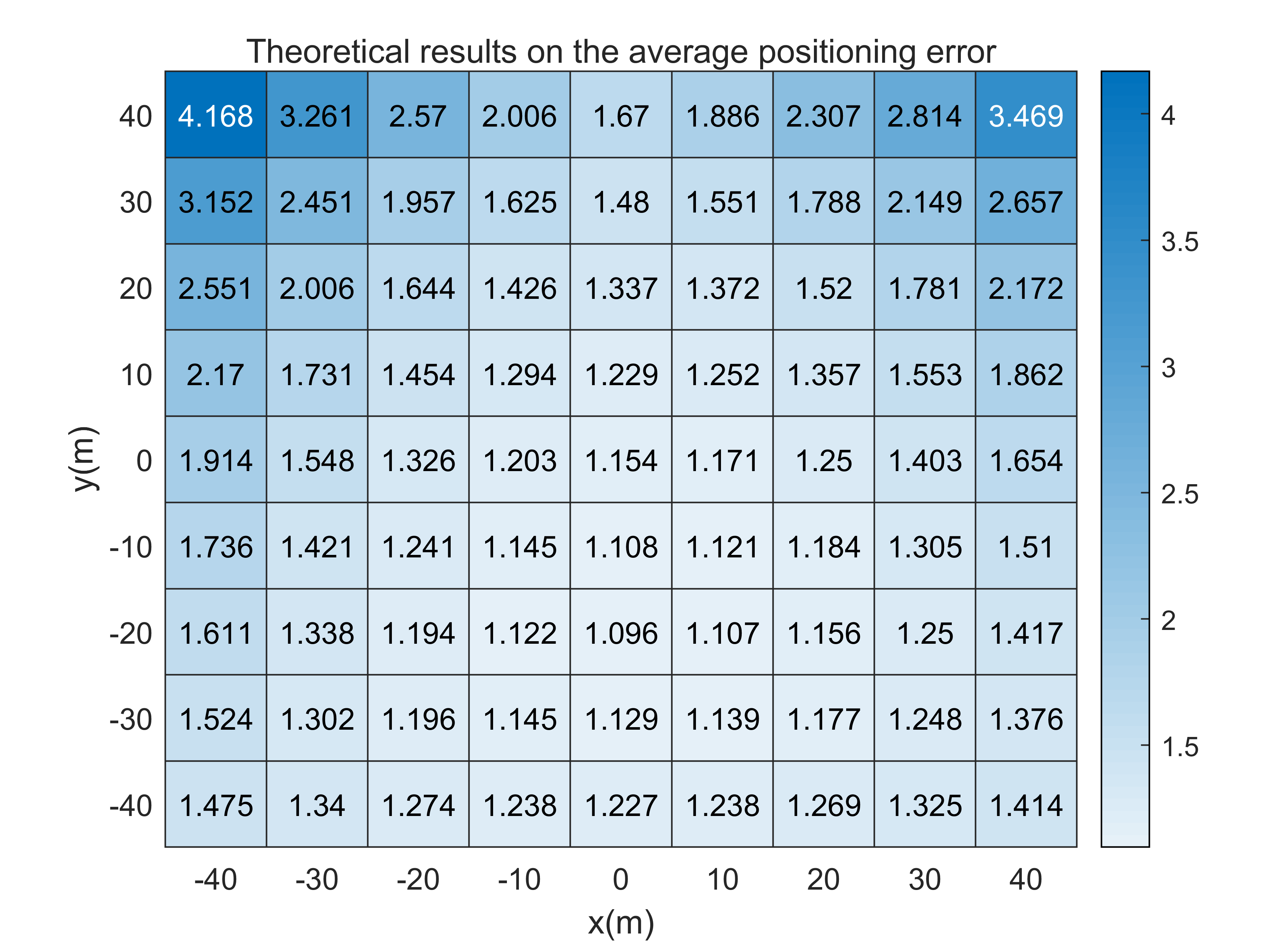}
	\caption{$\ $Theoretical average positioning error with $\sigma_{clock}=0 $.} \label{d_std_pic_secErr0_calcu}
	%    	\vspace{-1.0cm}%%压缩图片后间隔
\end{figure}

We first discuss an ideal case where complete clock synchronization is achieved among the three transmitters  with $\sigma_{clock}=0 $. Fig. \ref{d_std_pic_secErr0_calcu} shows the theoretical average positioning error in Eq. \eqref{trace}.

\begin{figure}[htbp]
	\centering
	\includegraphics[width=3.0in]{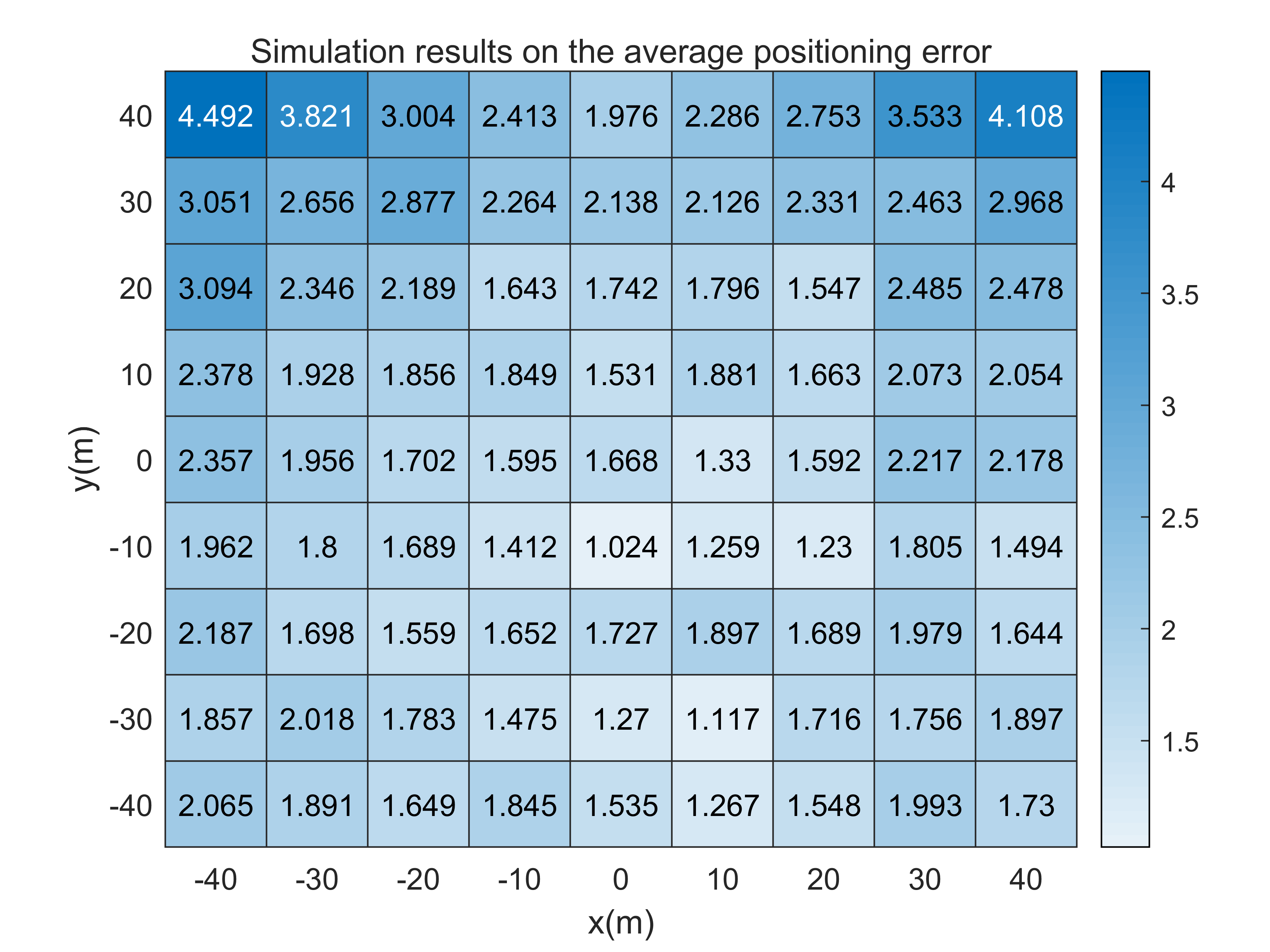}
	\caption{$\ $Simulated average positioning error with $\sigma_{clock}=0 $.} \label{d_std_pic_secErr0}
	\vspace{-0.3cm}%%压缩图片后间隔
\end{figure}

In the simulation with $\sigma_{clock}=0 $, the simulated positioning error is shown in Fig. \ref{d_std_pic_secErr0}, which is in good agreement with the theoretical error in Fig. \ref{d_std_pic_secErr0_calcu}. Lower positioning error can be observed in the center of location range.

\begin{figure}[htbp]
	\centering
	\includegraphics[width=3.0in]{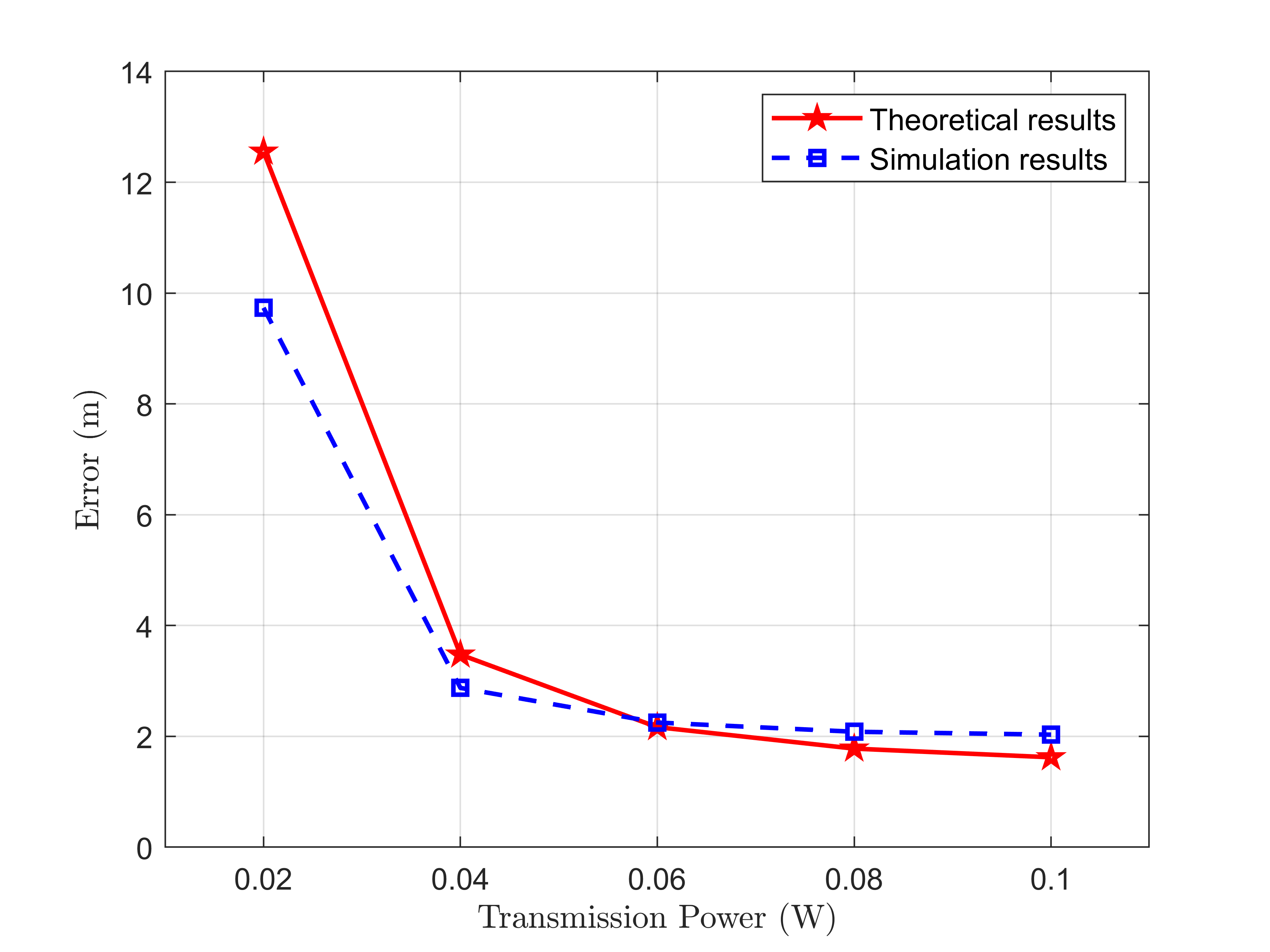}
	\caption{$\ $Theoretical and simulation results on the average positioning error with $\sigma_{clock}=0 $.} \label{d_std_mean_pi}
	\vspace{-0.3cm}%%压缩图片后间隔
\end{figure}

In order to explore the effect of transmission power on the positioning error, we carry out more simulations and compare the results with the theoretical one, as shown in Fig. \ref{d_std_mean_pi}. The horizontal axis represents the transmission power and vertical axis represents the average positioning error of $81$ receiver points in the location range in Fig. \ref{LocationRange}. It can be found that the positioning accuracy can be improved as the transmission power increases. However, as the transmission power increases to a certain extent, the accuracy improvement becomes saturated. This is due to the limitation of the PMT's sampling rate at the receiver, where the synchronization accuracy and positioning accuracy cannot be further improved via increasing the transmission power.
\subsection{Actual Case: Non-negligible Timing Error at Different Transmitters}
In the actual positioning system realization, a challenge is to consider the synchronization bias between different transmitters. After receiving the satellite PPS (pulse per second) signal at each transmitter, a square wave rising edge with a frequency of $10MHz$ is adopted to control the transmission of the synchronous sequence. The error of $10MHz$ square wave superimposed by the error of 1 PPS signal can be considered to meet the uniform distribution $U[0ns,100ns]$ according to Section \ref{timesec}.

\begin{figure}[htbp]
	\centering
	\includegraphics[width=3.0in]{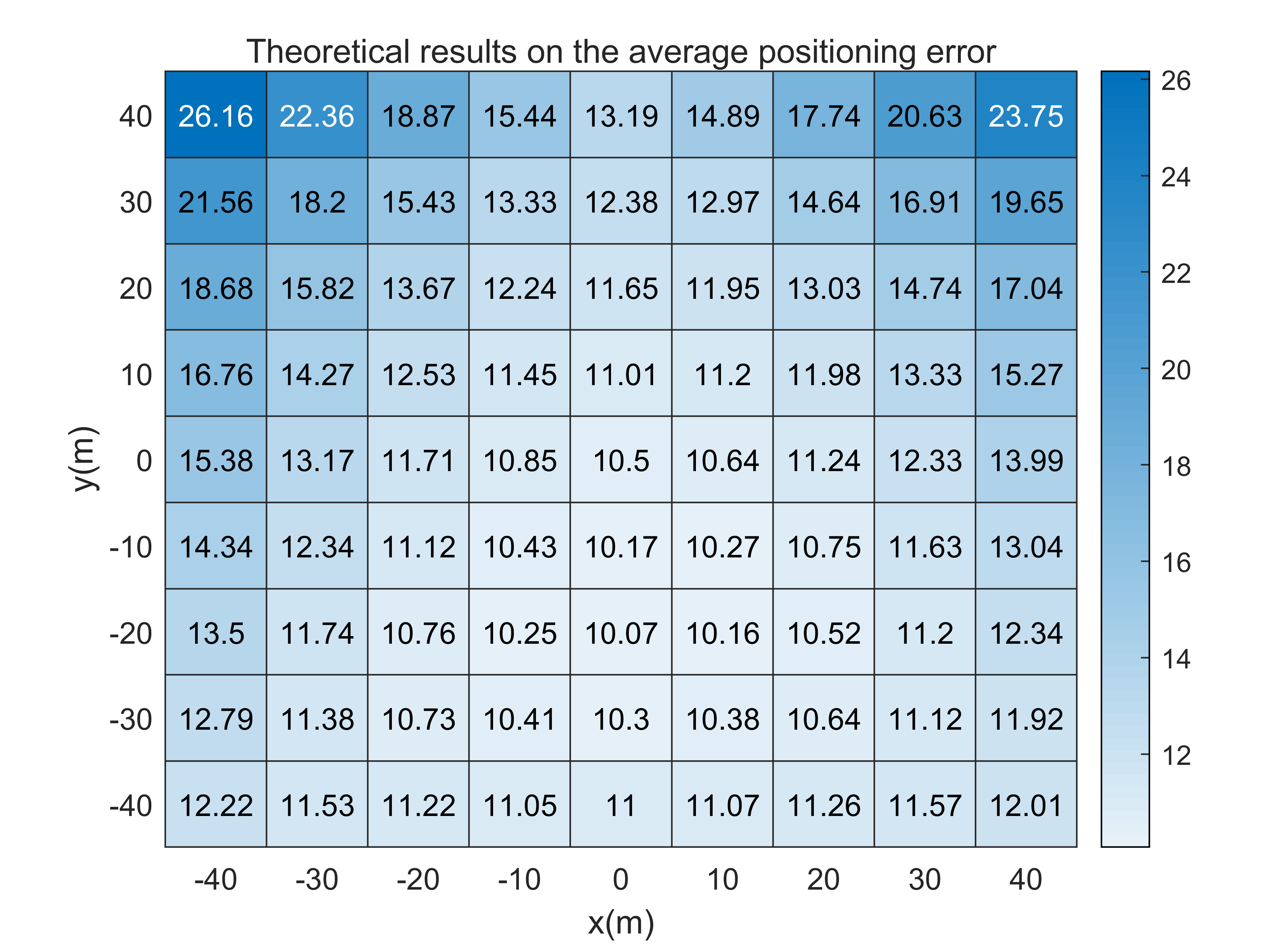}
	\caption{$\ $Theoretical average positioning error with non-negligible rising-edge bias.} \label{d_std_pic_secErr50_calcu}
	\vspace{-0.3cm}%%压缩图片后间隔
\end{figure}

In the simulation with timing bias between different transmitters, we assume a uniformly distributed random variable satisfying $U[0ns,100ns]$ on sending time difference of each transmitter under hardware realization. The positioning range is consistent with that in Subsection \ref{ssec.ideal}. Both theoretical and simulation results show that the transmitter-side timing error has a significant impact on the positioning accuracy. Note that there exists a triangle with three transmitters as the vertices. Inside the triangle, the simulated positioning error shows agreement with the theoretical one, all around $10$ meters. Outside the triangle, the positioning accuracy and the agreement between the theoretical accuracy and simulated one becomes degraded, primarily due to the accuracy of the first-order approximation in the theoretical derivation.
\begin{figure}[htbp]
	\centering
	\includegraphics[width=3.0in]{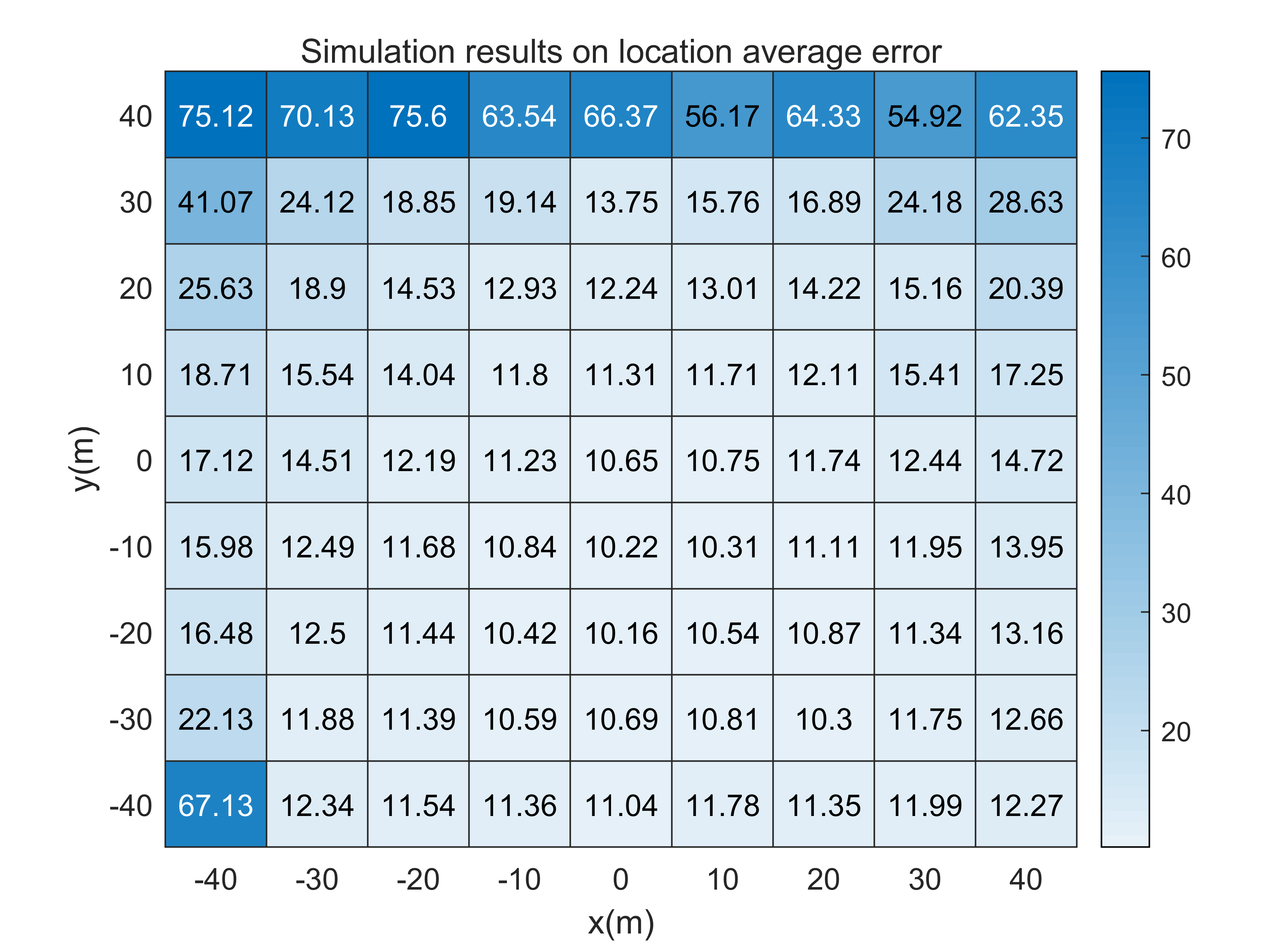}
	\caption{$\ $Simulated average positioning error with non-negligible rising-edge bias.} \label{d_std_pic_secErr50}
	%    	\vspace{-0.3cm}%%压缩图片后间隔
\end{figure}
\begin{figure}[htbp]
	\centering
	\includegraphics[width=3.0in]{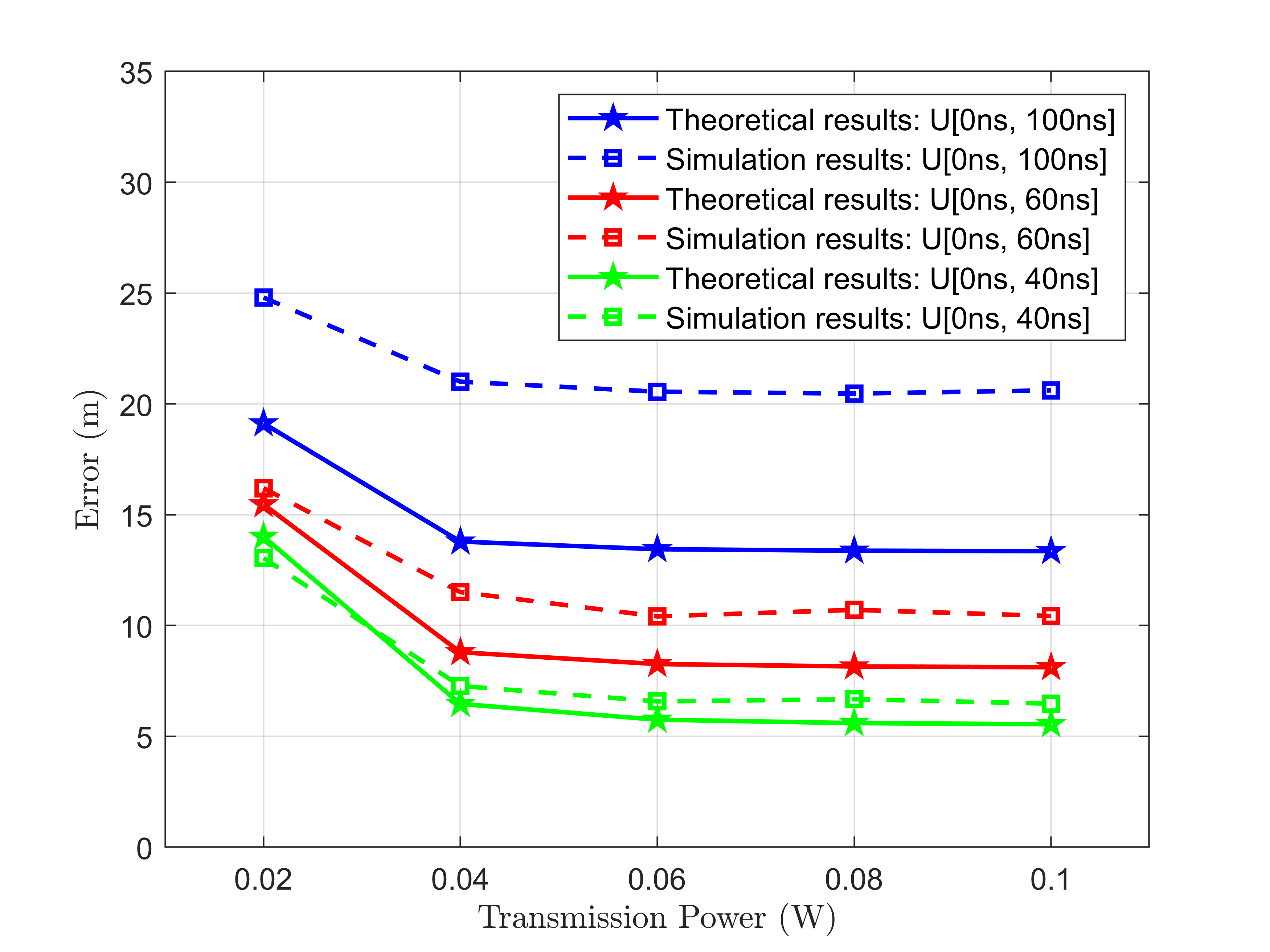}
	\caption{$\ $Theoretical and simulated average positioning error under transmitter timing error.} \label{d_std_mean_pi50}
	\vspace{-0.3cm}%%压缩图片后间隔
\end{figure}

\begin{figure}[htbp]
	\centering
	\includegraphics[width=3.0in]{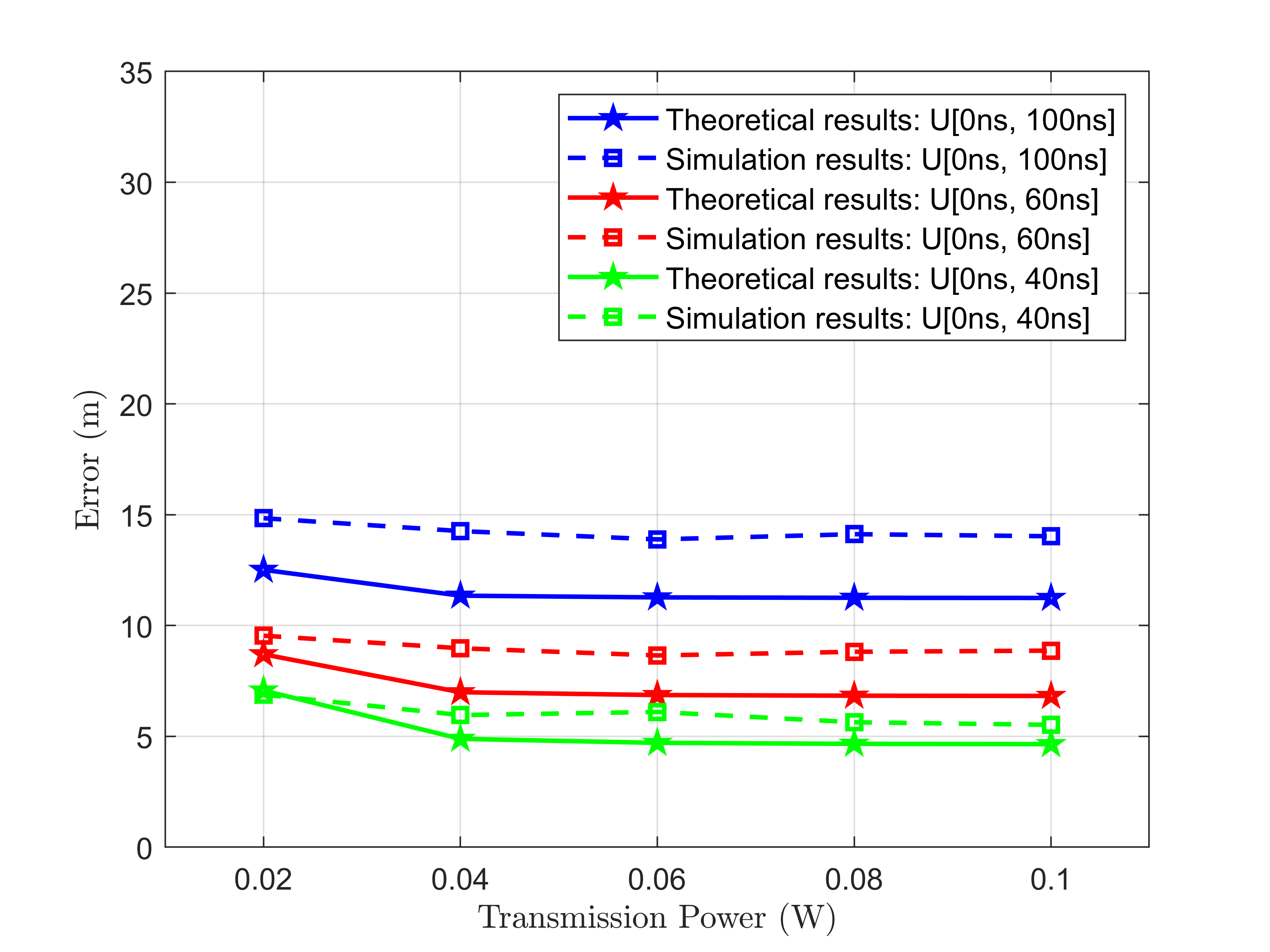}
	\caption{$\ $Theoretical and simulated average positioning error inside the triangle under transmitter timing error.} \label{d_std_mean_pi2}
	\vspace{-0.3cm}%%压缩图片后间隔
\end{figure}

In order to explore the influence of transmission power on the positioning accuracy with timing bias, we also conduct more simulations and compare the results with the theoretical ones, as shown in Fig. \ref{d_std_mean_pi50}. Both theoretical and simulated values are obtained assuming uniform distribution of misalignment in the transmitter clock rising edge under different parameters.

Similarly, the positioning error decreases with the transmission power. However, when the transmission power increases to a certain extent, the positioning error reduction becomes saturated. This is not only due to the limitation of the PMT's sampling rate at the receiver but also due to the atomic clock error satisfying uniform distribution at the transmitters, where the latter is the main reason of the positioning accuracy degradation compared with those shown in Fig. \ref{d_std_mean_pi} in Section \ref{ssec.ideal}. In the simulation where the rising edge error satisfies $U[0ns,100ns]$, the points in the region outside the triangle shows bigger gap between the simulated average positioning error and the theoretical one. When the positioning range is inside the triangle as shown in Fig. \ref{d_std_mean_pi2}, the gap decreases significantly. Through the comparison of theoretical and simulated results, a larger rising edge misalignment of the transmitter clock leads to a larger positioning error. Such results imply that the positioning accuracy can be improved via more accurate clock alignment.

\section{Outdoor Positioning Experiment} \label{Outdoor Location Experiment}
%In outdoor experiments, we execute the time division transmission that A, B and C send synchronization sequences, which length satisfies $L=256$, in turn at fixed time interval $T=300us$. The symbol rate of transmission is $ 1Mbps$.
%%\begin{figure}[htbp]
%%	\centering
%%	\includegraphics[width=5.5in]{expLoc}
%%	\caption{$\ $Schematic diagram of location in outdoor experiment.} \label{expLoc}
%%	%    	\vspace{-1.0cm}%%压缩图片后间隔
%%\end{figure}
%The outdoor real-time experiment was carried out two times in total. Lay the three transmitters and the receiver R as Fig. \ref{TDOA} shows and the process of the outdoor experiment follows Fig. \ref{expPro}. R is placed in the triangle formed by the lines between one transmitter and another. 

\subsection{Hardware Realization of the Positioning System}
\begin{figure*}[htbp]
	\centering
	\includegraphics[width=4.5in]{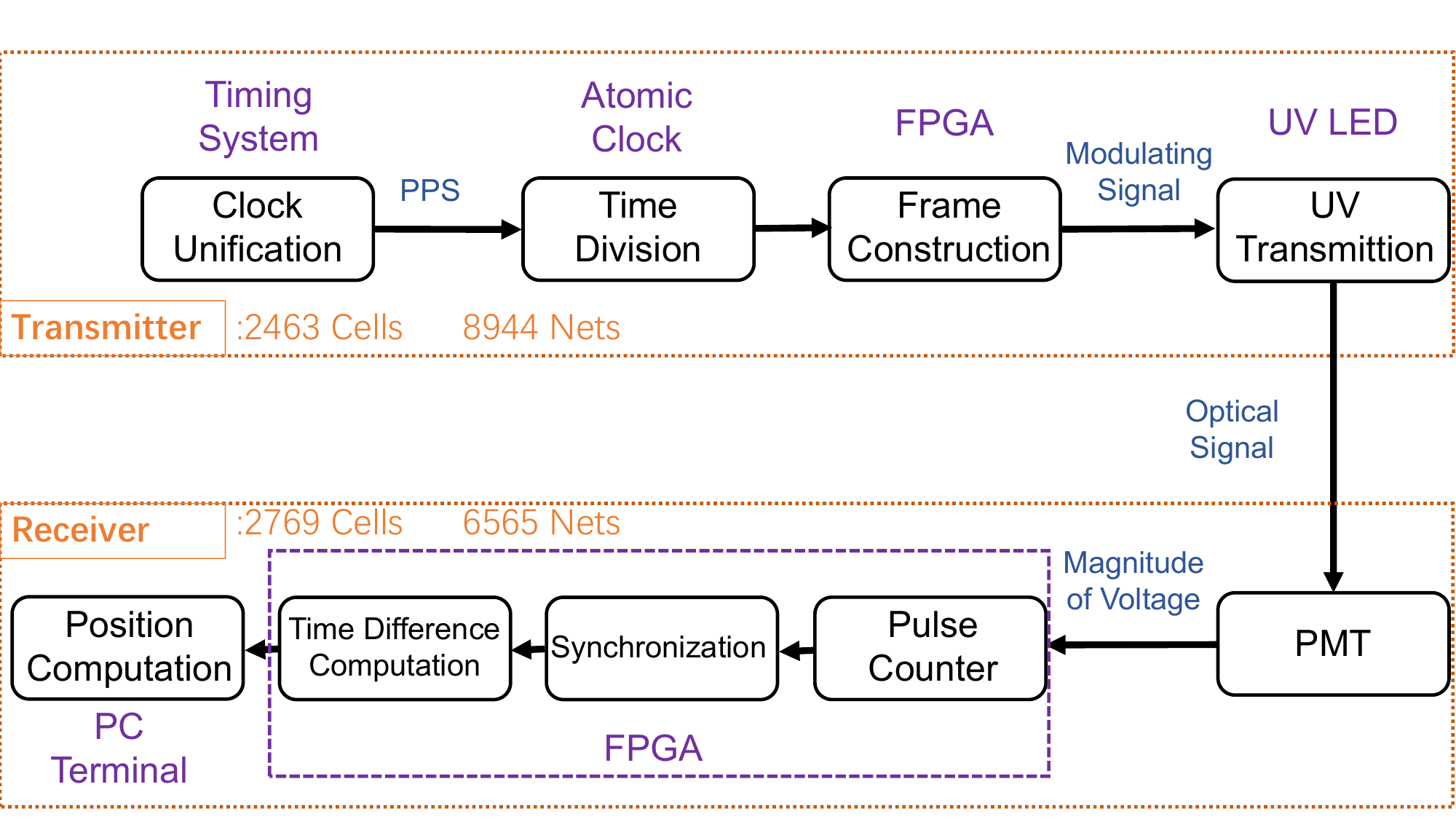}
	\caption{$\ $The functional block diagram of UV communication system.} \label{expPro}
	\vspace{-0.2cm}%%压缩图片后间隔
\end{figure*}

\begin{figure*}[htbp]
	\centering
	\includegraphics[width=4.5in]{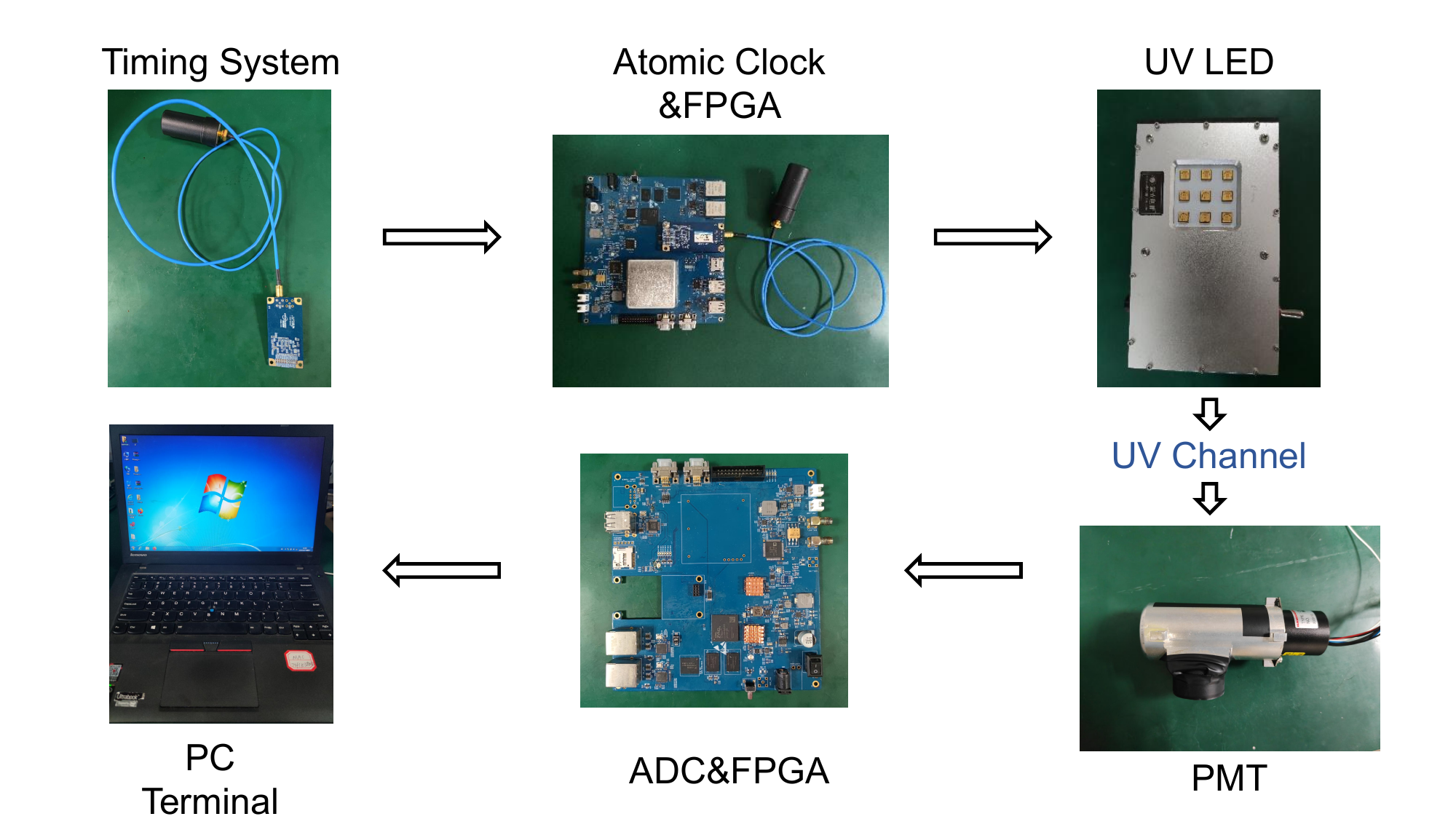}
	\caption{$\ $The hardware realization blocks for the UV positioning system.} \label{hardBlo}
	%    	\vspace{-1.0cm}%%压缩图片后间隔
\end{figure*}

The system block diagram and experimental system hardware realization are shown in Fig. \ref{expPro} and Fig. \ref{hardBlo}, respectively.

\begin{figure*}
	\centering
	\includegraphics[width=4.5in]{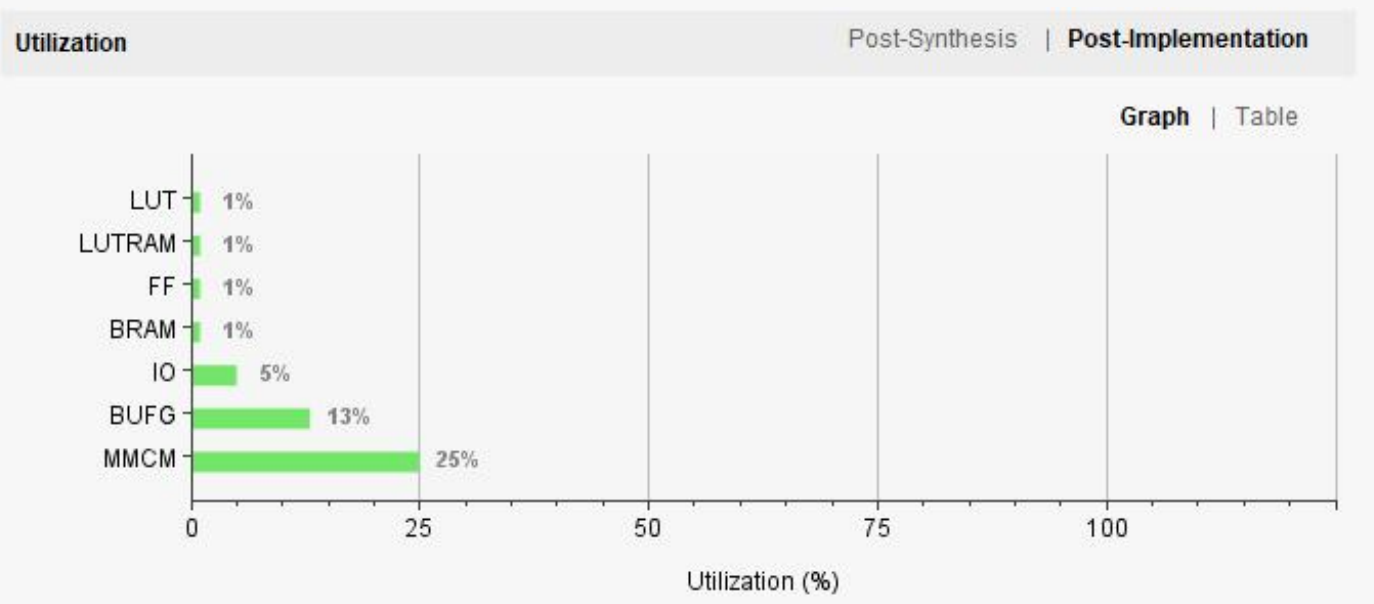}
	\caption{The resource utilization after hardware implementation at the transmitter side.}
	\label{utiTran}
	\vspace{-0.2cm}%%压缩图片后间隔
\end{figure*}

\begin{figure*}
	\centering
	\includegraphics[width=4.5in]{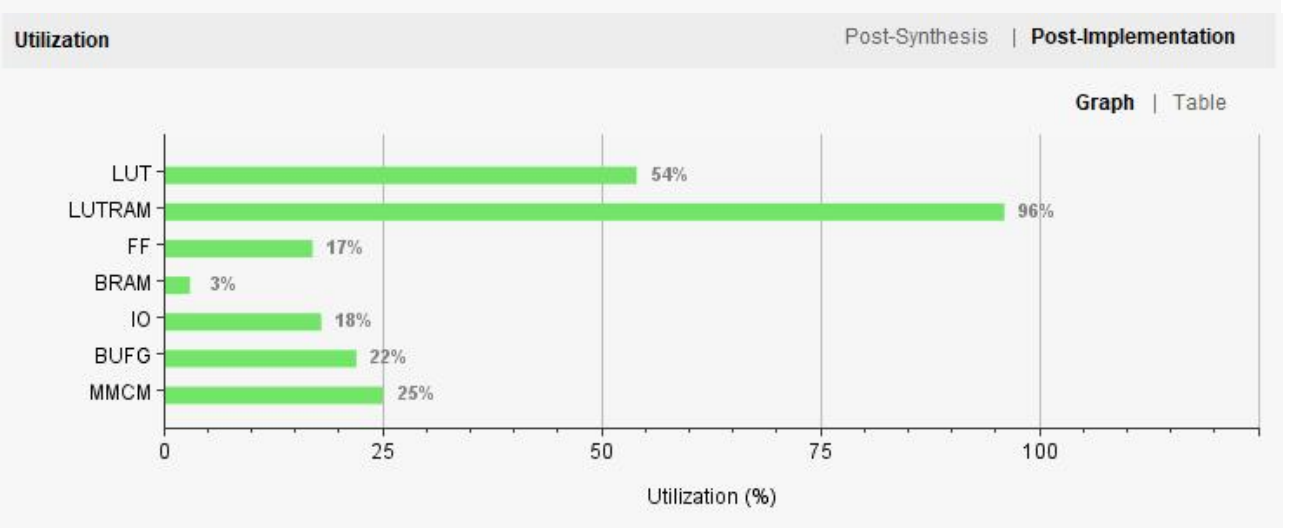}
	\caption{The resource utilization after hardware implementation at the receiver side.}
	\label{uti}
	\vspace{-0.2cm}%%压缩图片后间隔
\end{figure*}

At the transmitter side, the timing system including an antenna (unicorecomm, UT4B0) send Pulse Per Second (PPS) to Field Programmable Gate Array (FPGA) to control the time-division transmission by the atomic clock (CHENGDU ELECSPN, XHTF1033$\_$CPT), whose timing error satisfies uniform distribution $U[0ns,100ns]$ according to Section \ref{timesec}. The UV LED (266nm) transmits ultraviolet pilot signal modulated by FPGA. The resource utilization after hardware implementation at the transmitter side is shown in Fig. \ref{utiTran}.

At the receiver side, PMTs are employed as the photon-detector, which can convert the received
photoelectrons to analog pulses such that photon counting can be realized via pulse counting. 
HAMAMATSU PMT (R7154 module) detector is sealed into a shielding box, which is integrated
with an UV optical filter that passes the light signal of wavelength around 266 nm and blocks
the solar background radiation on other wavelengths. Analog-to-digital converters (ADCs) are employed such that the rising edge detection can be performed in the digital domain. The digital signal processing of synchronization and computation of the arrival time differences between receiving signals from different transmitters are performed sequentially on the FPGA board. The resource utilization after hardware implementation at the receiver side is shown in Fig. \ref{uti}. Then, the arrival time differences are sent to the receiver-side personal computer (RX-PC) through serial ports to obtain the real-time positioning results according to Eq. \eqref{lequ}. 

The corresponding real-time hardware realization is shown in Fig. \ref{hardBlo}. An atomic clock including an antenna is welded in the FPGA board and the timing system is connected to the FPGA board through pin interfaces. The UV LED consists of an array of $3\times3$ UV light beads. The spectral response of PMT ranges from $160nm$ to $320nm$ and the detection bandwith is wider than $200MHz$. The ADC sampling line connects the PMT to the ADC on the FPGA board, which outputs the processed data to the PC terminal.

We carried out a preliminary test including twelve receiver points on a small lawn, which is denoted as Experiment $\Rmnum{1}$ as shown in Table \ref{table.outdoor1}. The specific position coordinates and error results are presented. Points A, B and C indicate three transmitters, whose coordinates are shown in Table \ref{table.outdoor1}. Three transmitters A, B and C send synchronization sequences in a time-division manner with length $L=256$ and fixed time interval $T=300us$. The symbol rate of transmission is $ 1Mbps$ and the transmission power is $150mW$. The positioning error for a receiver point is calculated as the average error in ten estimates based on measurements. Theoretical error denotes the corresponding theoretical positioning error. The experimental average positioning error over the twelve points is $10.2130m$, while the corresponding theoretical average positioning error is $11.1303m$.

\subsection{Outdoor Test on the Playground}\label{testp}

%
%Weather: Sunny
%
%Air temperature: $\qty{25}{\degreeCelsius} \sim \qty{34}{\degreeCelsius}$
%
%Air Quality Index (AQI): $27$

We conducted an outdoor experiment in the university playground. The test was completed on July 23, 2023. The weather was sunny, with temperatures ranging from 25$^{\circ}$C to 34$^{\circ}$C, and the air quality index (AQI) was 27. The hardware and experimental processing are the same as those in the preliminary test. The transmission power is $150mW$.
\begin{figure}[htbp]
	\centering
	\includegraphics[width=3.0in]{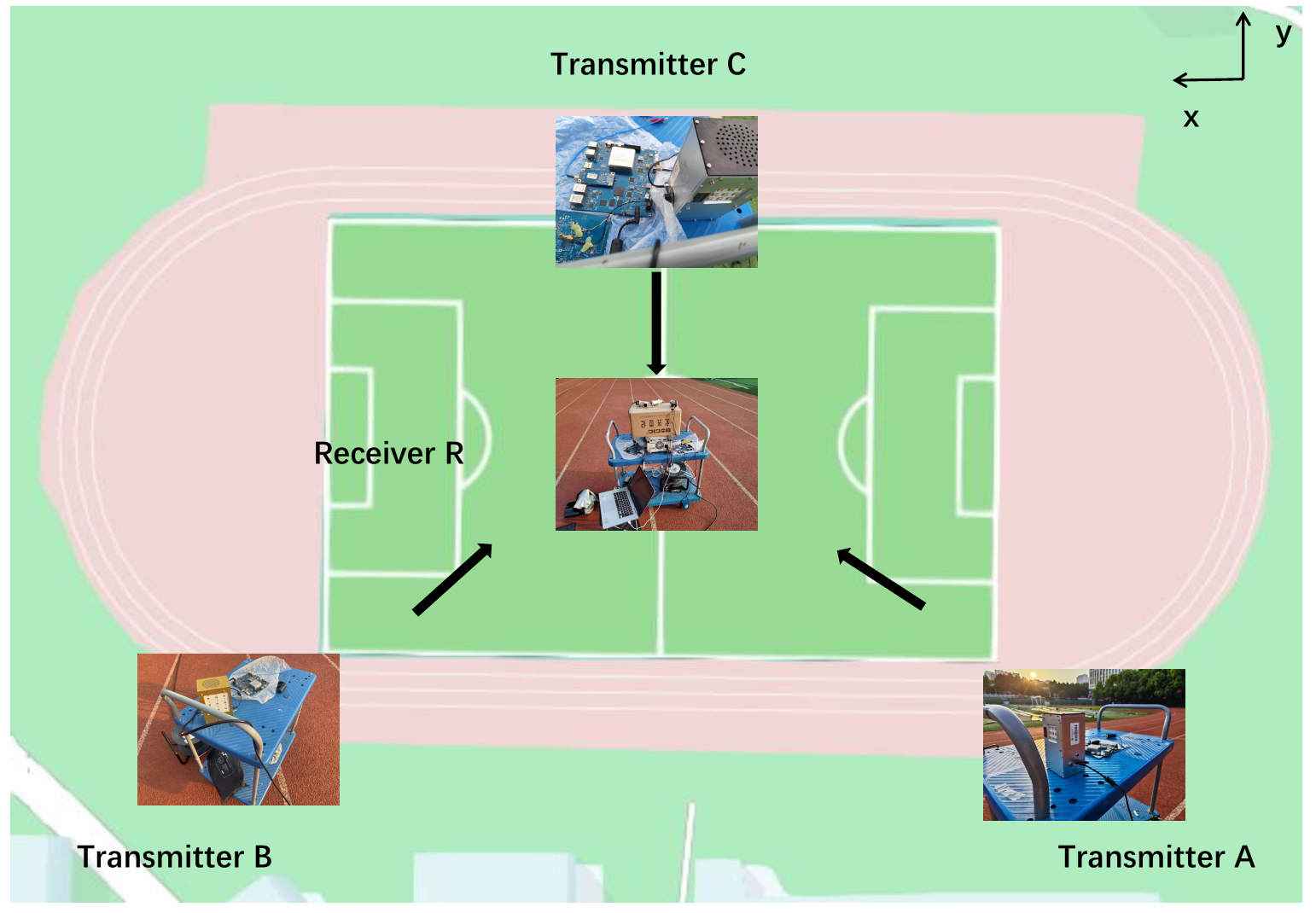}
	\caption{$\ $The positions of three transmitters and receiver $R$ in the outdoor experiment on the university playground.} \label{playground}
	%    	\vspace{-1.0cm}%%压缩图片后间隔
\end{figure}
%The outdoor real-time experiment was carried out two times in total. 
Lay the three transmitters and the receiver R as shown in Fig. \ref{playground}. The hardware processing of outdoor experiment follows Fig. \ref{expPro}. R is placed in the triangle connected by the three transmitters.

\begin{figure*}[htbp]
	\centering
	\includegraphics[width=4.2in]{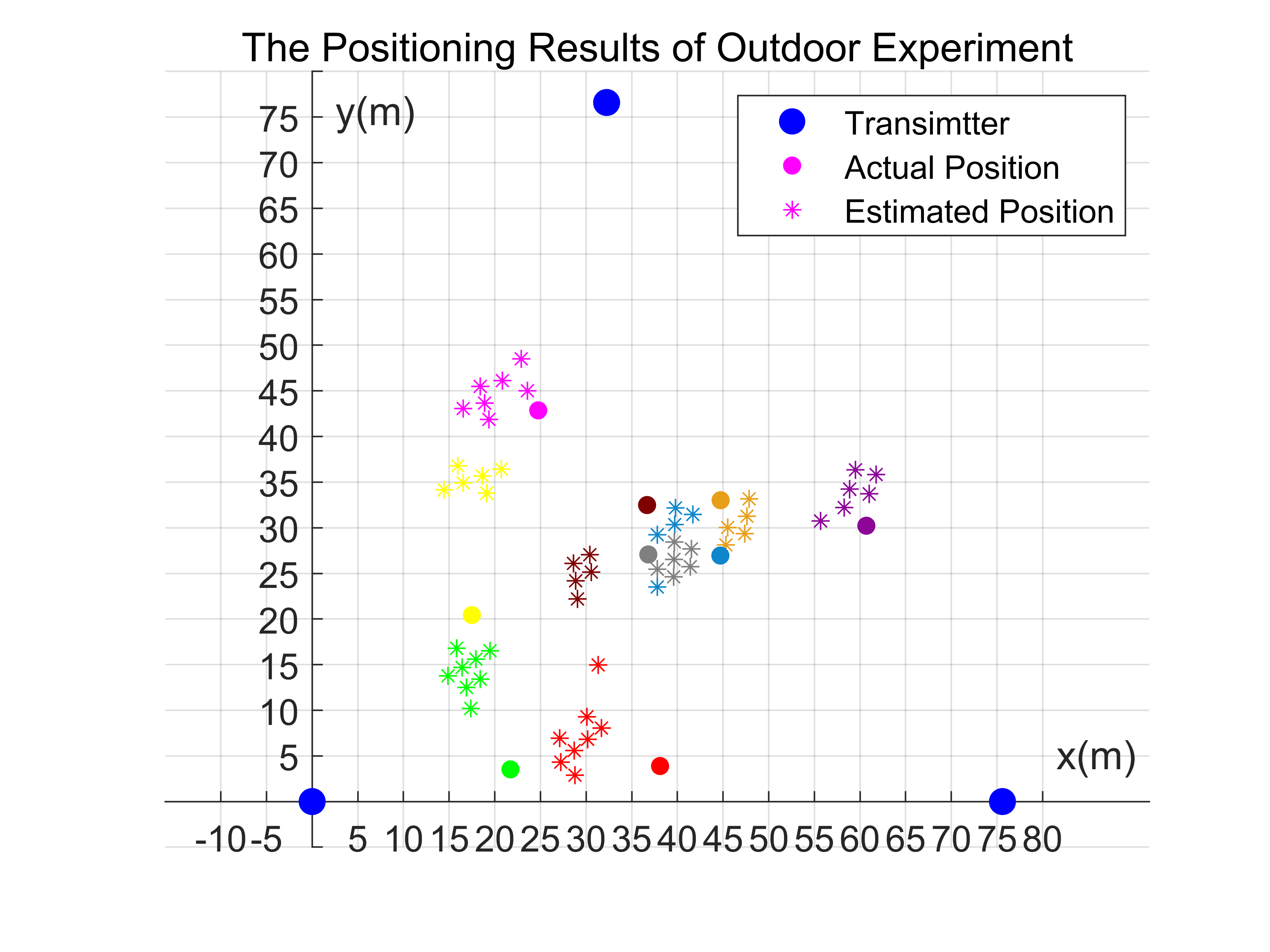}
	\caption{$\ $The positioning results in the outdoor experiment on the university playground.} \label{positionRst}
	%    	\vspace{-1.0cm}%%压缩图片后间隔
\end{figure*}

The positioning results are shown in Fig. \ref{positionRst}, where the three transmitters are marked as three big blue circles at three known locations. Each solid colored dot denotes the real receiver position, and the asterisk mark of the corresponding color denotes the positioning estimate of the receiver. The positioning performances at nine positions are tested, where ten estimates are obtained for each positioning. Clustering effect of the estimated position can be observed with small internal variance. The average distance from the ten estimated values to the center of the ten positions can reach the level around $2.1159m$. This is due to the fact that the positioning error mainly comes from the misalignment in the rising edge of the transmitter clock in the short time duration.

As shown in Table. \ref{table.outdoor1}, Experiment $\Rmnum{2}$ denotes the outdoor test including nine receiver points in total on the playground. The experimental average positioning error over the all nine points is $11.2000m$, while the corresponding theoretical average positioning error is $10.6725m$.
\subsection{Outdoor Test on a Grand Lawn}

%Date: 10/26/2023
%
%Weather: Cloudy
%
%Air temperature: $\qty{14}{\degreeCelsius} \sim \qty{28}{\degreeCelsius}$
%
%Air Quality Index (AQI): $90$

\begin{figure*}[htbp]
	\centering
	\includegraphics[width=4.5in]{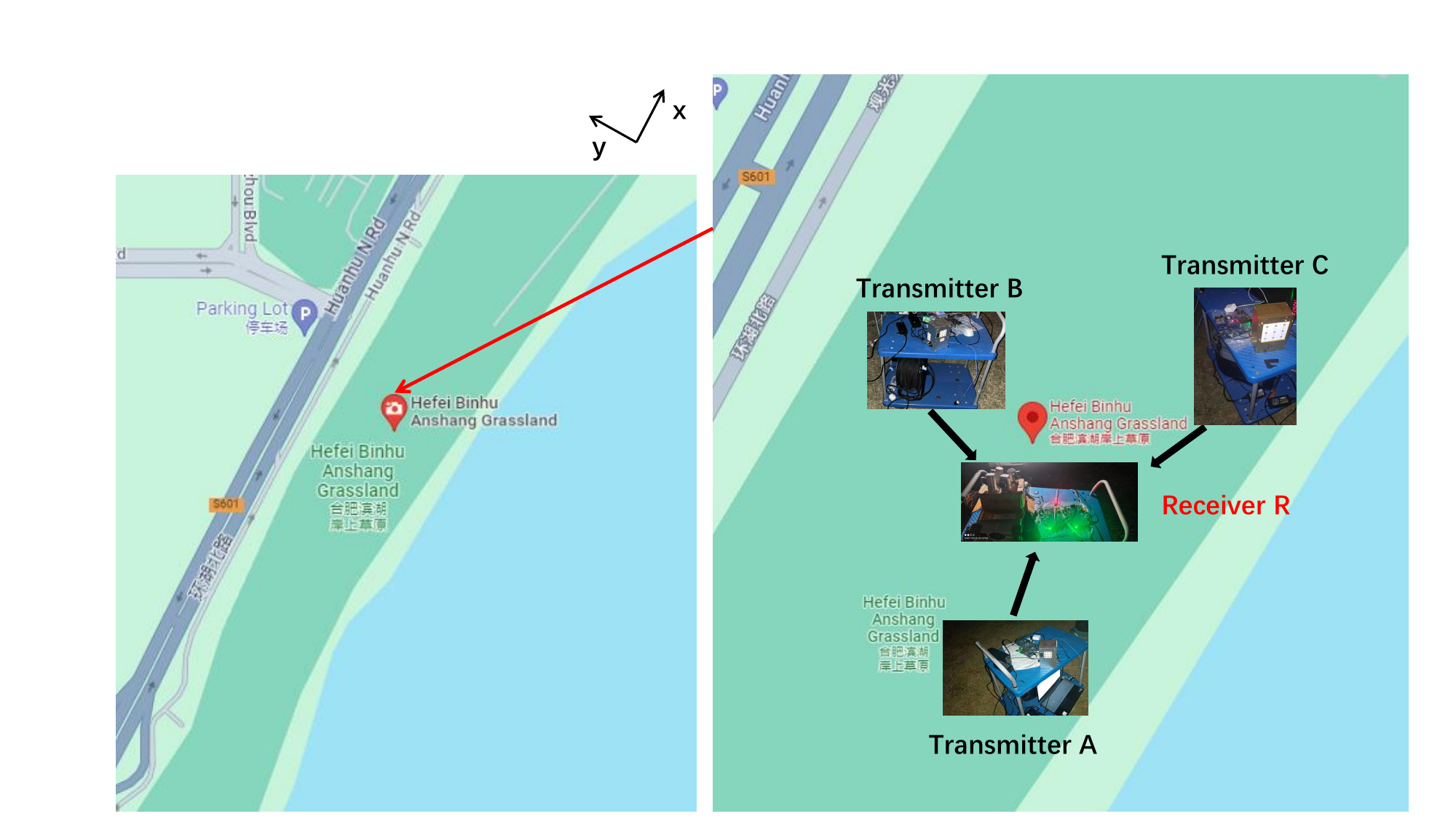}
	\caption{$\ $The positions of three transmitters and receiver $R$ in the outdoor experiment on the larger grand lawn.} \label{lawn}
	%    	\vspace{-1.0cm}%%压缩图片后间隔
\end{figure*}
\begin{figure*}[htbp]
	\centering
	\includegraphics[width=4.2in]{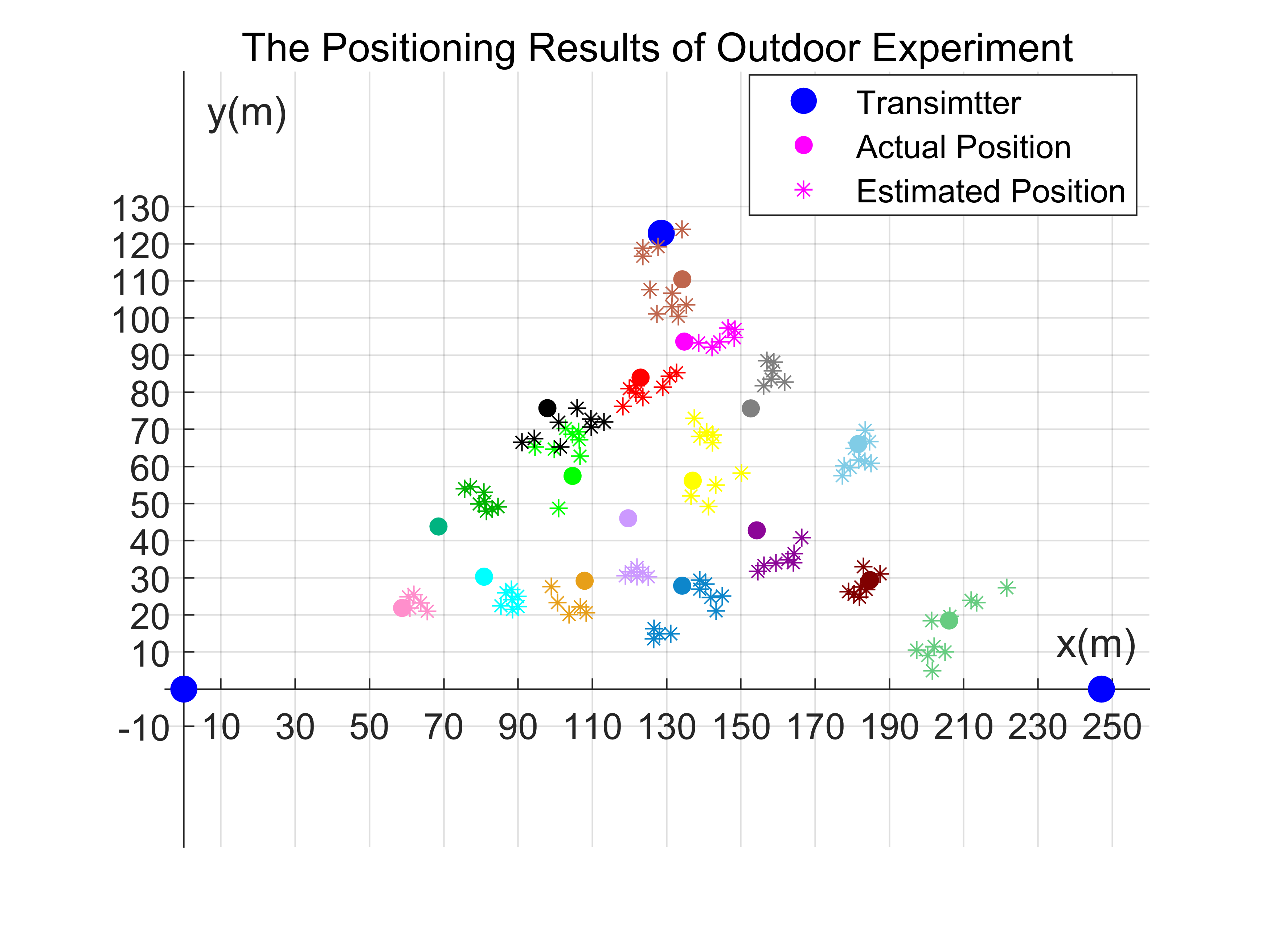}
	\caption{$\ $The positioning results in the outdoor experiment on the grand lawn.} \label{positionRst2}
	%   	\vspace{-0.5cm}%%压缩图片后间隔
\end{figure*}

We further conducted another outdoor test on a larger grand lawn on Hefei Binhu Anshang Grassland, where the layout of three transmitters and receiver R is shown in Fig. \ref{lawn}. The test was completed on October 26, 2023. The weather was cloudy, with temperatures ranging from 14$^{\circ}$C to 28$^{\circ}$C, and the air quality index (AQI) was 90. Receiver R is placed in the triangle connected by the three transmitters. The positioning at seventeen positions are tested, where ten estimates are obtained for each positioning. The hardware and experimental processing are the same as those in Section \ref{testp}. The transmission power is adjusted to $1W$.

The positioning results are shown in Fig. \ref{positionRst2}. The clustering effect of the estimated position near the actual position can also be observed. However, compared with Fig. \ref{positionRst}, the internal variance of the cluster is larger. We compute the average distance from the ten estimated values to the center of mass of the estimated cluster, where the average over seventeen receiver points reaches $5.0780m$. At some points, two clusters can be observed. This is due to wider distribution of the three transmitters, which tends to cause variations of misalignment. 

As shown in Table \ref{table.outdoor1}, the experimental average positioning error over the seventeen points in Experiment $\Rmnum{3}$ is $9.8310m$, while the corresponding theoretical average positioning error is $11.4894m$. Such results validate the consistency of outdoor experiments with the theoretical average positioning error. 

\begin{table*}[h]
	\begin{center}
		\renewcommand\arraystretch{2}
		\captionof{table}{The positioning accuracy in the three scenarios of outdoor experiments (unit m).}
		\label{table.outdoor1}
		\begin{tabular}{| c | c | c | c | c | c | c |}
			\hline
			Index &  $A$ & $B$ & $C$   & Experimental Error & Theoretical Error \\ \hline

			$\Rmnum{1}$ &$\left(30.2,53.9\right)$ & $\left(0,0\right)$& $\left(60.7,0\right)$ & 
			$ 10.2130 $& $ 11.1303 $     \\ \hline
			
			$\Rmnum{2}$ &$\left(0,0\right)$ & $\left(75.6,0\right)$ & $\left(32.2,76.6\right)$ 
			& $ 11.2000 $ & $ 10.6725 $ \\ \hline
			
			$\Rmnum{3}$ &$\left(0,0\right)$ & $\left(128.6,122.8\right)$& $\left(247.1,0\right)$ & 
			$ 9.8310 $& $ 11.4894 $     \\ \hline
		\end{tabular}
		%		\vspace{-3em} 
	\end{center}
\end{table*}

\subsection{Further Accuracy Improvement Roadmap}
%The specific position coordinates and error results are presented in Table. \ref{table.outdoor1}. A, B and C indicate three transmitters. The table shows their coordinates. Location error for a receiver point is the average of the errors in the ten estimates during the short time for positioning measurements. Theoretical error is the corresponding fitting result. Experiment $\Rmnum{2}$ denotes the outdoor test including nine receiver points in total on the playground. The average positioning error of the all nine points in experiment $\Rmnum{2}$ is $11.2000m$, while the corresponding theoretical fit average is $10.6725m$. Experiment $\Rmnum{3}$ including seventeen receiver points is carried out on a grand lawn as Fig. \ref{lawn} shows. Similarly, the average positioning error of the seventeen points in experiment $\Rmnum{3}$ is $9.8310m$, while the corresponding theoretical fit average is $11.4894m$. The numerical consistency indicates that the results of outdoor experiments verify the accuracy of the theoretical fitting values. It is worth noting that when the error in the rising edge of the transmitter clock is reduced, the positioning accuracy will be significantly improved.

In addition to changing the transmitter-side atomic clock with that with smaller error in the rising edge, the positioning performance can also be improved via time synchronization processing between the three transmitters. We simulate a virtual differential positioning, where the timing error estimate is randomly selected from all ten estimates for each receiver point to obtain A-B clock correction and B-C clock correction. After the correction, the average positioning error over all nine points in Experiment $\Rmnum{2}$ can reach $3.1269m$; and the average positioning error over seventeen points in Experiment $\Rmnum{3}$ can reach $7.4675m$. This is due to the observation that in Experiment III, the $10$ estimates are sometimes divided into more than one cluster, such that the positioning error reduction is smaller than that of Experiment $\Rmnum{2}$. Such results imply that if the clock system misalignment can be estimated and transmitted to the receiver, the positioning accuracy can potentially be significantly improved.

%The location error is $2.9867m$, while its corresponding theoretical fitted error is $ 2.3277m $ and simulation error is $2.4512m $. Consistency among the three error results shows the reliability of time division transimission scheme in hardware implementation and practical application. 

%\begin{figure}[htbp]
%	\centering
%	\includegraphics[width=5.5in]{exp_err_rst}
%	\caption{$\ $Results of positioning error in outdoor experiments.} \label{expErr}
%	%    	\vspace{-1.0cm}%%压缩图片后间隔
%\end{figure}

%\begin{table*}[h]
%	\begin{center}
	%		\renewcommand\arraystretch{2}
	%		\captionof{table}{The location results for outdoor experiment.}
	%		\label{table.outdoor1}
	%		\begin{tabular}{| c | c | c | c | c | c | c |}
		%			\hline
		%			$r_{13}$ &$\hat{r}_{13}$ & $r_{23}$ & $\hat{r}_{23}$  & $ R $ & $ \hat{R} $  & Location error \\ \hline
		%		$-7.6m$ &$-12m$ & $2.98m$& $0m$ & $ \left(32.12m,13.32m\right) $ & $ \left(33.7200m,15.8419m\right) $  & $ 2.9867m$ \\ 
		%			\hline
		%		\end{tabular}
	%%		\vspace{-3em} 
	%	\end{center}
%\end{table*}
%\input{chapters/Sec6_SeqOptimization}
%\input{chapters/Sec7_Simulation}
\section{Conclusion} \label{section.conclusion}
We have designed, prototyped and tested the TDOA positioning system via UV transmission. Based on the synchronization accuracy, the theoretical positioning error is analyzed, which shows satisfactorily agreement with the simulated positioning error. We have also conducted hardware realization as well as three outdoor tests. The average positioning errors obtained from the experiments show agreement with the theoretical results. The positioning error is dominated by the timing synchronization issue of the three transmitters, and can be further reduced via time synchronization among three transmitters.
%\input{Appendices}

%\begin{footnotesize}
%	\bibliographystyle{IEEEtran}
%	\bibliography{bib/Synchronization_Sequence}
%\end{footnotesize}

\small{\baselineskip = 10pt
	\bibliographystyle{IEEEtran}
	\bibliography{Synchronization_Sequence}

\begin{thebibliography}{10}
\providecommand{\url}[1]{#1}
\csname url@samestyle\endcsname
\providecommand{\newblock}{\relax}
\providecommand{\bibinfo}[2]{#2}
\providecommand{\BIBentrySTDinterwordspacing}{\spaceskip=0pt\relax}
\providecommand{\BIBentryALTinterwordstretchfactor}{4}
\providecommand{\BIBentryALTinterwordspacing}{\spaceskip=\fontdimen2\font plus
\BIBentryALTinterwordstretchfactor\fontdimen3\font minus
  \fontdimen4\font\relax}
\providecommand{\BIBforeignlanguage}[2]{{%
\expandafter\ifx\csname l@#1\endcsname\relax
\typeout{** WARNING: IEEEtran.bst: No hyphenation pattern has been}%
\typeout{** loaded for the language `#1'. Using the pattern for}%
\typeout{** the default language instead.}%
\else
\language=\csname l@#1\endcsname
\fi
#2}}
\providecommand{\BIBdecl}{\relax}
\BIBdecl

\bibitem{301830}
Y.~Chan and K.~Ho, ``A simple and efficient estimator for hyperbolic
  location,'' \emph{IEEE Transactions on Signal Processing}, vol.~42, no.~8,
  pp. 1905--1915, 1994.

\bibitem{6151186}
K.~C. Ho, ``Bias reduction for an explicit solution of source localization
  using tdoa,'' \emph{IEEE Transactions on Signal Processing}, vol.~60, no.~5,
  pp. 2101--2114, 2012.

\bibitem{259534}
K.~Ho and Y.~Chan, ``Solution and performance analysis of geolocation by
  tdoa,'' \emph{IEEE Transactions on Aerospace and Electronic Systems},
  vol.~29, no.~4, pp. 1311--1322, 1993.

\bibitem{4682575}
N.~Liu, Z.~Xu, and B.~M. Sadler, ``Low-complexity hyperbolic source
  localization with a linear sensor array,'' \emph{IEEE Signal Processing
  Letters}, vol.~15, pp. 865--868, 2008.

\bibitem{5778025}
H.~Xu and Y.~Wang, ``A linear algorithm based on tdoa technique for uwb
  localization,'' in \emph{2011 International Conference on Electric
  Information and Control Engineering}, 2011, pp. 1013--1015.

\bibitem{1323254}
K.~Ho and W.~Xu, ``An accurate algebraic solution for moving source location
  using tdoa and fdoa measurements,'' \emph{IEEE Transactions on Signal
  Processing}, vol.~52, no.~9, pp. 2453--2463, 2004.

\bibitem{6151850}
N.~Liu, Z.~Xu, and B.~M. Sadler, ``Geolocation performance with biased range
  measurements,'' \emph{IEEE Transactions on Signal Processing}, vol.~60,
  no.~5, pp. 2315--2329, 2012.

\bibitem{6131130}
S.-Y. Jung, S.~Hann, and C.-S. Park, ``Tdoa-based optical wireless indoor
  localization using led ceiling lamps,'' \emph{IEEE Transactions on Consumer
  Electronics}, vol.~57, no.~4, pp. 1592--1597, 2011.

\bibitem{6844864}
M.~A. Khalighi and M.~Uysal, ``Survey on free space optical communication: A
  communication theory perspective,'' \emph{IEEE Communications Surveys \&
  Tutorials}, vol.~16, no.~4, pp. 2231--2258, 2014.

\bibitem{1159099}
M.-C. Jeong, J.-S. Lee, S.-Y. Kim, S.-W. Namgung, J.-H. Lee, M.-Y. Cho, S.-W.
  Huh, Y.-S. Ahn, J.-W. Cho, and J.-S. Lee, ``8 x 10-gb/s terrestrial optical
  free-space transmission over 3.4 km using an optical repeater,'' \emph{IEEE
  Photonics Technology Letters}, vol.~15, no.~1, pp. 171--173, 2003.

\bibitem{5175684}
E.~Ciaramella, Y.~Arimoto, G.~Contestabile, M.~Presi, A.~D'Errico, V.~Guarino,
  and M.~Matsumoto, ``1.28-tb/s (32 $\times$ 40 gb/s) free-space optical wdm
  transmission system,'' \emph{IEEE Photonics Technology Letters}, vol.~21,
  no.~16, pp. 1121--1123, 2009.

\bibitem{6165319}
W.~J. Miniscalco and S.~A. Lane, ``Optical space-time division multiple
  access,'' \emph{IEEE/OSA Journal of Lightwave Technology}, vol.~30, no.~11,
  pp. 1771--1785, 2012.

\bibitem{6458971}
S.~Zhang, S.~Watson, J.~J.~D. McKendry, D.~Massoubre, A.~Cogman, E.~Gu, R.~K.
  Henderson, A.~E. Kelly, and M.~D. Dawson, ``1.5 gbit/s multi-channel visible
  light communications using cmos-controlled gan-based leds,'' \emph{IEEE/OSA
  Journal of Lightwave Technology}, vol.~31, no.~8, pp. 1211--1216, 2013.

\bibitem{4063386}
V.~W.~S. Chan, ``Free-space optical communications,'' \emph{IEEE/OSA Journal of
  Lightwave Technology}, vol.~24, no.~12, pp. 4750--4762, 2006.

\bibitem{938713}
H.~Willebrand and B.~Ghuman, ``Fiber optics without fiber,'' \emph{IEEE
  Spectrum}, vol.~38, no.~8, pp. 40--45, 2001.

\bibitem{1299334}
D.~Kedar and S.~Arnon, ``Urban optical wireless communication networks: the
  main challenges and possible solutions,'' \emph{IEEE Communications
  Magazine}, vol.~42, no.~5, pp. S2--S7, 2004.

\bibitem{1527982}
G.~Shaw, A.~Siegel, and J.~Model, ``Ultraviolet comm links for distributed
  sensor networks,'' in \emph{Digest of the LEOS Summer Topical Meetings,
  2005.}, 2005, pp. 39--40.

\bibitem{8641355}
A.~Vavoulas, H.~G. Sandalidis, N.~D. Chatzidiamantis, Z.~Xu, and G.~K.
  Karagiannidis, ``A survey on ultraviolet c-band (uv-c) communications,''
  \emph{IEEE Communications Surveys Tutorials}, vol.~21, no.~3, pp. 2111--2133,
  2019.

\bibitem{10.1117/12.582002}
\BIBentryALTinterwordspacing
D.~M. Reilly, D.~T. Moriarty, and J.~A. Maynard, ``{Unique properties of solar
  blind ultraviolet communication systems for unattended ground sensor
  networks},'' in \emph{Unmanned/Unattended Sensors and Sensor Networks}, E.~M.
  Carapezza, Ed., vol. 5611, International Society for Optics and
  Photonics.\hskip 1em plus 0.5em minus 0.4em\relax SPIE, 2004, pp. 244 -- 254.
  [Online]. Available: \url{https://doi.org/10.1117/12.582002}
\BIBentrySTDinterwordspacing

\bibitem{5342313}
H.~Ding, G.~Chen, A.~K. Majumdar, B.~M. Sadler, and Z.~Xu, ``Modeling of
  non-line-of-sight ultraviolet scattering channels for communication,''
  \emph{IEEE Journal on Selected Areas in Communications}, vol.~27, no.~9, pp.
  1535--1544, 2009.

\bibitem{Xiao2011NonlineofsightUS}
H.~Xiao, Y.~Zuo, J.~Wu, H.~Guo, and J.~Lin, ``Non-line-of-sight ultraviolet
  single-scatter propagation model.'' \emph{Optics Express}, vol. 19 18, pp.
  17\,864--75, 2011.

\bibitem{6923957}
G.~Chen, L.~Liao, Z.~Li, R.~J. Drost, and B.~M. Sadler, ``Experimental and
  simulated evaluation of long distance nlos uv communication,'' in \emph{2014
  9th International Symposium on Communication Systems, Networks Digital Sign
  (CSNDSP)}, 2014, pp. 904--909.

\bibitem{Liao2015LongdistanceNU}
L.~Liao, R.~J. Drost, Z.~Li, T.~Lang, B.~M. Sadler, and G.~Chen,
  ``Long-distance non-line-of-sight ultraviolet communication channel analysis:
  experimentation and modelling,'' \emph{IET Optoelectronics}, vol.~9, pp.
  223--231, 2015.

\bibitem{Borah2021SingleAD}
D.~K. Borah, V.~R. Mareddy, and D.~G. Voelz, ``Single and double scattering
  event analysis for ultraviolet communication channels,'' \emph{Optics
  Express}, vol. 29 4, pp. 5327--5342, 2021.

\bibitem{9295804}
Z.~Shen, J.~Ma, T.~Shan, and P.~Su, ``Improved monte carlo integration models
  for ultraviolet communications,'' in \emph{2020 IEEE 20th International
  Conference on Communication Technology (ICCT)}, 2020, pp. 168--172.

\bibitem{6877718}
C.~Gong and Z.~Xu, ``Non-line of sight optical wireless relaying with the
  photon counting receiver: A count-and-forward protocol,'' \emph{IEEE
  Transactions on Wireless Communications}, vol.~14, no.~1, pp. 376--388, 2015.

\bibitem{8275030}
C.~Gong, K.~Wang, Z.~Xu, and X.~Wang, ``On full-duplex relaying for optical
  wireless scattering communication with on-off keying modulation,'' \emph{IEEE
  Transactions on Wireless Communications}, vol.~17, no.~4, pp. 2525--2538,
  2018.

\bibitem{7830279}
M.~H. Ardakani, A.~R. Heidarpour, and M.~Uysal, ``Performance analysis of
  relay-assisted nlos ultraviolet communications over turbulence channels,''
  \emph{IEEE/OSA Journal of Optical Communications and Networking}, vol.~9,
  no.~1, pp. 109--118, 2017.

\bibitem{5599260}
N.~D. Chatzidiamantis, G.~K. Karagiannidis, and M.~Uysal, ``Generalized
  maximum-likelihood sequence detection for photon-counting free space optical
  systems,'' \emph{IEEE Transactions on Communications}, vol.~58, no.~12, pp.
  3381--3385, 2010.

\bibitem{7112175}
C.~Gong and Z.~Xu, ``Lmmse simo receiver for short-range non-line-of-sight
  scattering communication,'' \emph{IEEE Transactions on Wireless
  Communications}, vol.~14, no.~10, pp. 5338--5349, 2015.

\bibitem{7047703}
M.~A. El-Shimy and S.~Hranilovic, ``Spatial-diversity imaging receivers for
  non-line-of-sight solar-blind uv communications,'' \emph{IEEE/OSA Journal of
  Lightwave Technology}, vol.~33, no.~11, pp. 2246--2255, 2015.

\bibitem{8332484}
G.~Wang, K.~Wang, C.~Gong, D.~Zou, Z.~Jiang, and Z.~Xu, ``A 1mbps real-time
  nlos uv scattering communication system with receiver diversity over 1km,''
  \emph{IEEE Photonics Journal}, vol.~10, no.~2, pp. 1--13, 2018.

\bibitem{2022arXiv220801559Y}
S.~{Yu}, C.~{Gong}, and Z.~{Xu}, ``{The design and optimization of
  synchronization sequence for Ultraviolet communication},'' \emph{arXiv
  e-prints}, p. arXiv:2208.01559, Aug. 2022.

\bibitem{8322671}
Q.~Zou, W.~Xia, Y.~Zhu, J.~Zhang, B.~Huang, F.~Yan, and L.~Shen, ``A vlc and
  imu integration indoor positioning algorithm with weighted unscented kalman
  filter,'' in \emph{2017 3rd IEEE International Conference on Computer and
  Communications (ICCC)}, 2017, pp. 887--891.

\bibitem{8121297}
W.~Tang, J.~Zhang, B.~Chen, Y.~Liu, Y.~Zuo, S.~Liu, and Y.~Dai, ``Analysis of
  indoor vlc positioning system with multiple reflections,'' in \emph{2017 16th
  International Conference on Optical Communications and Networks (ICOCN)},
  2017, pp. 1--3.

\bibitem{7394267}
T.~The~Son, H.~Le-Minh, F.~Mousa, Z.~Ghassemlooy, and N.~Van~Tuan, ``Adaptive
  correction model for indoor mimo vlc using positioning technique with node
  knowledge,'' in \emph{2015 International Conference on Communications,
  Management and Telecommunications (ComManTel)}, 2015, pp. 94--98.

\bibitem{8533834}
B.~Ghimire, J.~Seitz, and C.~Mutschler, ``Indoor positioning using ofdm-based
  visible light communication system,'' in \emph{2018 International Conference
  on Indoor Positioning and Indoor Navigation (IPIN)}, 2018, pp. 1--8.

\bibitem{7859314}
B.~Lin, X.~Tang, Z.~Ghassemlooy, C.~Lin, and Y.~Li, ``Experimental
  demonstration of an indoor vlc positioning system based on ofdma,''
  \emph{IEEE Photonics Journal}, vol.~9, no.~2, pp. 1--9, 2017.

\bibitem{7928991}
F.~Akhoundi, A.~Minoofar, and J.~A. Salehi, ``Underwater positioning system
  based on cellular underwater wireless optical cdma networks,'' in \emph{2017
  26th Wireless and Optical Communication Conference (WOCC)}, 2017, pp. 1--3.

\bibitem{10092387}
Y.~Zhang, Z.~Wei, Z.~Liu, C.~Cheng, Z.~Wang, X.~Tang, Y.~Yang, C.~Yu, and H.~Y.
  Fu, ``Optical communication and positioning convergence for flexible
  underwater wireless sensor network,'' \emph{IEEE/OSA Journal of Lightwave
  Technology}, vol.~41, no.~16, pp. 5321--5327, 2023.

\bibitem{7275471}
Y.~Nakazawa, H.~Makino, K.~Nishimori, D.~Wakatsuki, and H.~Komagata,
  ``Led-tracking and id-estimation for indoor positioning using visible light
  communication,'' in \emph{2014 International Conference on Indoor Positioning
  and Indoor Navigation (IPIN)}, 2014, pp. 87--94.

\bibitem{9158795}
J.-H. Kim and S.~Moon, ``Recursive bayesian estimation based indoor fire
  location by fusing rotary uv sensors,'' in \emph{2020 IEEE/ASME International
  Conference on Advanced Intelligent Mechatronics (AIM)}, 2020, pp. 528--533.

\end{thebibliography}
}

\end{document}